\documentclass[useAMS,usenatbib]{mn2e}
\usepackage{amssymb}
\usepackage{graphicx}
\DeclareGraphicsExtensions{.jpg,.pdf,.png,.eps,.pdf}
\usepackage{url}
\usepackage{color}
\usepackage{tabulary}
\usepackage{multirow}
\usepackage{amsmath}
\usepackage{amsfonts}
\usepackage{ulem}
\usepackage[referable]{threeparttablex}
\usepackage{booktabs, dcolumn}

\graphicspath{ {pictures/} }
\newcommand{\beq}{\begin{equation}}
\newcommand{\eeq}{\end{equation}}
\newcommand{\simlt}{\mathrel{\hbox{\rlap{\hbox{\lower4pt\hbox{$\sim$}}}\hbox{$<$}}}}
\newcommand{\simgt}{\mathrel{\hbox{\rlap{\hbox{\lower4pt\hbox{$\sim$}}}\hbox{$>$}}}}

\newcommand{\msun}{\;\mathrm{M}_{\odot}}
\newcommand{\rsun}{\;\mathrm{R}_{\odot}}
\newcommand{\zsun}{\;\mathrm{Z}_{\odot}}
\newcommand{\gpy}{\;\mathrm{Gpc^{-3}~yr^{-1}}}

\title[BH-BH mergers from Pop III stars]{On the likelihood of detecting
gravitational waves from Population III compact object binaries}

\author[K. Belczynski et al.]  {\parbox{\textwidth}{Krzysztof
    Belczynski$^{1}$\thanks{email: chrisbelczynski@gmail.com}, Taeho
    Ryu$^{2}$, Rosalba Perna$^{2}$, Emanuele Berti$^{3}$, \\ Takamitsu
    L. Tanaka$^{2}$, \vspace*{0.3cm}Tomasz Bulik$^{1,4}$}\\ 
$^{1}$Astronomical
  Observatory, Warsaw University, Ujazdowskie 4, 00-478 Warsaw,
  Poland\\ 
$^{2}$Department of Physics and Astronomy, Stony Brook
  University, Stony Brook, NY 11794-3800, USA\\ 
$^{3}$Department of
  Physics and Astronomy, The University of Mississippi, University, MS
  38677, USA\\ 
$^{4}$Instituto de Astronomía, Universidad Nacional Autónoma
  de México. Km 103 Carretera Tijuana-Ensenada,\\ \ 22860 Ensenada, Baja
  California, Mexico }

\begin{document}

\maketitle

\label{firstpage}

\begin{abstract}
  We study the contribution of binary black hole (BH-BH) mergers from
  the first, metal-free stars in the Universe (Pop~III) to
  gravitational wave detection rates. Our study combines initial
  conditions for the formation of Pop~III stars based on N-body
  simulations of binary formation (including rates, binary fraction,
  initial mass function, orbital separation and eccentricity
  distributions) with an updated model of stellar evolution specific
  for Pop~III stars. We find that the merger rate of these Pop~III
  BH-BH systems is relatively small ($\lesssim 0.1 \gpy$) at low
  redshifts ($z<2$), where it can be compared with the LIGO empirical
  estimate of $9$--$240\gpy$~\citep{LigoO1b}.  The predicted rates are
  even smaller for Pop~III double neutron star and black hole neutron
  star mergers. Our rates are compatible with those of \cite{Hartwig2016}, 
  but significantly smaller than those found in previous 
  work~\citep{Bond1984b,Belczynski2004,Kinugawa2014,Kinugawa2016b}.
  We explain the reasons for this discrepancy by means of detailed
  model comparisons and point out that 
  {\it (i)} identification of Pop~III BH-BH mergers may not be possible by 
            advanced LIGO, and  
  {\it (ii)} the level of stochastic gravitational wave background from
             Pop~III mergers may be lower than recently
             estimated~\citep{Kowalska2012,Inayoshi2016,Dvorkin2016}.
  We further estimate gravitational wave detection
  rates for third-generation interferometric detectors. Our
  calculations are relevant for low to moderately rotating Pop~III
  stars. We can now exclude significant ($>1$ per cent) contribution
  of these stars to low-redshift BH-BH mergers. However, it remains to
  be tested whether (and at what level) rapidly spinning Pop~III stars
  (homogeneous evolution) can contribute to BH-BH mergers in the local
  Universe.
\end{abstract}


\section{Introduction}
\label{sec:intro}

The discovery of gravitational waves by the LIGO collaboration
\citep{DiscoveryPaper}, in addition to probing gravity in an extreme
regime, has also opened a new window on the end state of massive
stars, and the compact objects that they leave behind. The first
detected event, GW150914, resulted from the merger of two massive
black holes (BH-BH) with chirp mass
$M_{\rm chirp}\equiv
(M_1M_2)^{3/5}/(M_1+M_2)^{1/5}=28^{+2}_{-2}\msun$, where $M_1$ and
$M_2$ are the individual source-frame BH masses. The fact that this
signal was observed so early during the first LIGO observing run (O1)
suggests that similar systems could be rather common. The chirp mass
$M_{\rm chirp}= 8.9^{+0.3}_{-0.3}\msun$ of the second detected event,
GW151226, was smaller \citep{LigoO1a}; a third lower-significance
trigger, LVT151012, if astrophysical in origin, was produced by a BH-BH 
binary with chirp mass in an intermediate range,
$M_{\rm chirp}=10^{+1}_{-1}\msun$ \citep{LigoO1b}.

While these few detections already show that there is a spread of
masses in the BH-BH mass distribution, it is the higher end of the
mass distribution which is particularly intriguing. In general,
high-mass BH-BH systems should form from very massive, metal-poor binary 
stars~\citep{Belczynski2010a}.
These stars are expected to be more numerous at higher redshifts, and
the requirement that they should be detectable within the LIGO horizon
sets some constraints on their merger time scales.

\begin{table*}
  \caption{Summary of the two models of Pop~III binary formation}  
 \centering
\begin{threeparttable}
  \setlength\extrarowheight{4pt}
	\begin{tabulary}{0.7\linewidth}{c| c c c c c c }
		\hline
		\hline
		model & 	number of stars $N$  per run& gas number density $n$ & 	initial spatial range & IMF\tnote{a} & 	$M_{\rm min}$ & 	$M_{\rm max}$ 	   \\
		\hline
		FS1& \multirow{2}{*}{5} & \multirow{2}{*}{$10^{6}~\mathrm{cm}^{-3}$ }& $2000~\mathrm{AU}$ & \multirow{2}{*}{$\alpha=0.17$ }& $0.1\msun$& 	$140\msun$        \\
		FS2 & 	       &    & $10-20~\mathrm{AU}$ & & $0.1\msun$ & 	$200\msun$       \\
		\hline
		\hline
	\end{tabulary}
	\label{tab:tab3}
	\begin{tablenotes}
		\item[a] The masses of the stars are drawn from a
              top-heavy IMF with $\alpha=0.17$
              ($\frac{dN}{dM}=M^{-\alpha}$,
              \citealt{StacyBromm13}). The distributions of the
              initial binary parameters used for the population
              synthesis calculations in Table \ref{tab.init} are
              based on those binaries which dynamically
              formed and were not subsequently disrupted by 
              dynamical interactions.
        \end{tablenotes}
\end{threeparttable}
\end{table*}

\citet{Belczynski2016b} reported a suite of numerical simulations of
BH-BH binary formation via the evolution of isolated binary stars. They
found that the progenitor stars of GW150914 had masses in the range of
$40-100\msun$, and formed in an environment where the metallicity is
less than 10 per cent of the solar metallicity.  Their progenitors
were likely formed when the Universe was about 2~Gyr old.
This, so called classical evolution channel, was also recently studied by
other authors (e.g., \citealt{Eldridge2016, Lipunov2017}).

Alternative scenarios for the formation of massive BH-BH binaries invoke
dynamical interactions at the center of star clusters
(e.g., \citealt{Ziosi2014, Rodriguez2016b, Askar2017}). For
example, single BHs can pair with a BH companion via three-body binary
formation and binary-mediated exchange interactions, which tend to
result in the ejection of the lightest BH, while the two most massive
BHs form a binary (e.g., \citealt{Heggie2003, Chatterjee2012, Morscher2015, 
Ryu+16, Chatterjee2016}).

Massive BH-BH formation was also proposed by binary evolution of rapidly
rotating stars, so called homogeneous evolution channel (e.g., \citealt{
Marchant2016, Mandel2016a, Woosley2016, Eldridge2016}).

The formation of massive BH-BH binaries would also be a natural outcome if
the BHs were the end products of Population~III (Pop~III) stars. These
are believed to be the first stars formed in the Universe, and hence
would naturally occur in metal-free environments
(e.g. \citealt{Omukai1998, Abel2002, Bromm2002}).  The possibility
that binaries of Pop~III star remnants could be contributing sources
of gravitational waves has been considered by a number of authors
\citep[see e.g.][]{Bond1984b,Belczynski2004,Kulczycki2006,Kinugawa2014,Hartwig2016},
who investigated a range of initial mass functions and initial binary
parameters.  In particular, \cite{Kinugawa2014} concluded that Pop~III
BH binary remnants can account for a significant fraction of current
and future gravitational wave detections.

In this paper we revisit this important question, motivated by the
fact that the computation of the local merger rate of Pop~III BH-BH 
binaries is very sensitive to the choice of the initial binary parameters, 
such as masses and initial orbital separations. We couple Pop~III initial
conditions determined via $N$-body simulations of binary formation
\citep{Ryu+16} of stars born in primordial halos 
\citep{Stacy+10,Greif+12} with a state-of-the-art numerical
computation of binary evolution (\citealt{Belczynski2016b}, with
updates specific to Pop~III evolution). We estimate the contribution
to the merger rates detectable by current and future detectors using
phenomenological models calibrated to numerical relativity simulations
of the binary BH merger signal, as in \cite{Dominik2015}. We predict
significantly lower merger rates with respect to \cite{Kinugawa2014},
and we discuss the underlying reasons for this discrepancy.

The paper is organized as follows. Sec.~2 describes the
(dynamically-determined) initial conditions for the Pop~III binary
stars. Sec.~3 details how they are evolved, as well as their redshift
distribution. The specific evolutionary scenarios ensuing from our
initial conditions are described in Sec.~4. In Sec.~5 we describe our
findings for the properties of the BH-BH population and we compute
gravitational wave detection rates for Advanced LIGO and several planned future
instruments. We devote Sec.~6 to a detailed model comparison with
\citet{Kinugawa2014}. In Sec.~7 we summarize our findings and indicate
possible directions for future work.

\section{Initial Properties of Pop~III stars}
\label{sec: InitialProperties}

In this study, we track the evolutions of Pop~III stars in binaries using
\texttt{StarTrack}. We use the models of \citet{Ryu+16} to determine 
the initial properties of the Pop~III binary stars---that is, {\it the mass of 
the primary star, the mass ratio between secondary and primary star, the 
semi-major axis and the eccentricity}. We briefly review their models 
below, but refer the reader to \citet{Ryu+16} for details. 

Using $N$-body simulations, \cite{Ryu+16} investigated the formation
of Pop~III X-ray binaries in star-forming gas clouds. They considered
multiple systems of Pop~III stars embedded in a uniform gas
medium, and included the physical effects from the gas medium 
(i.e. dynamical friction and background potential). Their
simulations followed the dynamics of the stars initially in
quasi-Keplerian orbit on a disk until isolated and stable binaries 
had formed.

For this study, we consider two specific scenarios with very different 
physical size of a gas cloud (mini-halo). This choice is motivated by the 
two available state-of-the-art Pop~III star formation numerical models in
mini-halos at high-redshift: large mini-halos; $\sim 2000\mathrm{AU}$ 
\citep{StacyBromm13} and small mini-halos; $10$--$20\mathrm{AU}$ \citep{Greif+12}.
These two models, while being very different, encompass the Pop~III star
formation uncertainties. In each scenario, a mini-halo is populated with 
$N=5$ single stars. They are placed at random positions within a given
mini-halo and are subjected to dynamical friction and are allowed to
dynamically interact with each other. As a result some of these stars form
binaries and occasionally higher-multiplicity systems. Such simulation is
repeated ($\sim 250$ times) to obtain initial distributions of Pop~III binary 
star parameters.

\begin{enumerate}
	\item \textit {Model FS1: moderate orbital separations}
	
          This model assumes that Pop~III stars form in a gas
          cloud of spatial range of $\sim2000~\mathrm{AU}$
          \citep{StacyBromm13}.  {The number density of the
            pristine gas medium is $10^{6}~\mathrm{cm}^{-3}$ and the
            masses of stars follow a top-heavy initial mass function
            (IMF) $\frac{dN}{dM}=M^{-\alpha}$ with $\alpha=0.17$,
            $M_{\text{max}}=140\msun$ and $M_{\text{min}} =0.1\msun$
            \citep{StacyBromm13}. The binary parameters are collected when 
            the binary begins to shrink predominantly via dynamical friction 
            with mini-halo gas after single stars and binaries (or triples) 
            are formed and isolated from one another; in other words, when no 
            further dynamical interactions between stars are expected (at 
          $t\sim1~\mathrm{Myr}$).}
	\vspace{0.05in}
	\item \textit {Model FS2: small orbital separations}
	
	In the second model we consider a rather small gas cloud of spatial 
	range of  $\sim 10 - 20~\mathrm{AU}$, motivated by the findings of \citet{Greif+12}. 
	Their simulations showed that as a result of fragmentation of gas clouds, 
	multiple, less massive protostars form around the most massive one, several 
        AU apart from each other. We adopt the same number density of the gas
	medium in a mini-halo as above ($10^{6}~\mathrm{cm}^{-3}$).
        In order to mimic their findings, we use the same IMF as in the FS1 model, 
        but alter $M_{\text{max}}$ in each of $5$ drawings. In the first
	drawing we adopt $M_{\text{max}}=200\msun$. Subsequently, we generate stars 
        with $M_{\text{max}}=200\msun - $ [the sum of the masses of the previously 
        generated stars]. In this model, the dynamical stellar interactions typically 
        end in a few to tens of thousands years due to smaller initial separations 
        between stars. The initial conditions for binary evolution with {\tt
	StarTrack} population synthesis are extracted at $t\sim1~\mathrm{Myr}$. 
	
\end{enumerate}

A summary of the main features of our models of Pop~III binary dynamical formation 
is given in Table~\ref{tab:tab3}. Note that in the simulations of both models
triples also are formed, but only the inner binaries in the triples were the ones 
taken into account. The distributions of parameters for binaries formed in the
above simulations are taken as initial input for population synthesis evolutionary 
calculations and are given in Table~\ref{tab.init}. 
This dynamical approach to determine the initial binary properties of Pop~III stars 
is an important difference with respect to previous studies.

\section{Evolution of Pop~III stars}
\label{model}

Population synthesis calculations were performed with the {\tt StarTrack} 
code \citep{Belczynski2002,Belczynski2008a}. Recently we updated this 
code with improved physics. The improvements relevant for massive star 
evolution include updates to the treatment of CE evolution \citep{Dominik2012},
 the compact object masses produced by core collapse/supernovae \citep{Fryer2012,Belczynski2012}, 
and observationally constrained star formation and metallicity evolution over 
cosmic time \citep{Belczynski2016b}. Here we discuss the existing updates and also 
introduce another set of updates that are especially relevant for Pop~III stars.

\subsection{Radius evolution}

In our model we employ modified \cite{Hurley2000} rapid evolutionary formulae. 
These formulae do not include the effects of stellar rotation, and they are 
limited in both metallicity ($Z=0.03 - 0.0001$) and initial star mass range 
($M_{\rm zams}=0.08 - 80\msun$). We have extended and calibrated the use of these 
formulae to higher mass ($M_{\rm zams}=0.08 - 150\msun$; \citealt{Belczynski2008a}).
The effects of stellar rotation on stellar evolution become significant for high 
rotation speeds: e.g., homogeneous evolution at rotation speeds close to breakup 
velocity is very different \citep{Marchant2016,Mandel2016a,Woosley2016} and it is 
not taken into account within our model. 

\begin{figure}
        \hspace*{-0.5cm}
        \includegraphics[width=9.1cm]{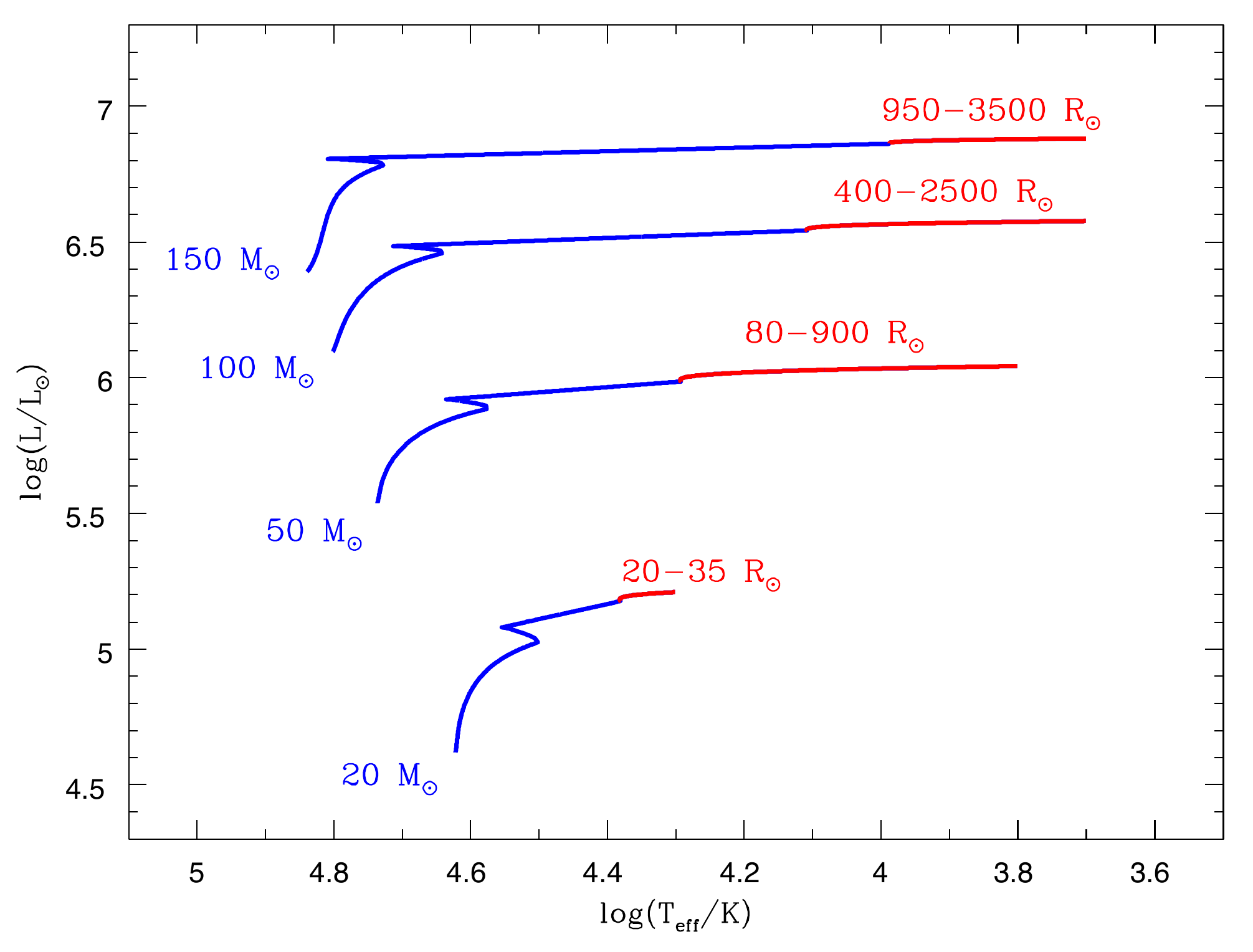} 
        \vspace*{-0.0cm}
	\caption{
		H-R diagram for massive stars: potential BH progenitors. We show tracks
		from ZAMS to CO core ignition. The blue part marks main sequence and
		Hertzsprung gap, while the red part marks the later evolutionary stages (core 
		He burning and beyond). Radial expansion during the later evolutionary stages
		is marked on each track; at this late evolutionary stages stars are cool
		enough to possibly develop deep convective envelopes and in case of Roche 
                Lobe overflow (RLOF) the common envelope (CE) evolution may follow. 
                At earlier evolutionary stages, RLOF either leads to stable mass transfer 
                (system survival) or CE that always ends up in a merger ending binary 
                evolution.  
	}
	\label{fig.radius}
\end{figure}

We use our lowest metallicity model ($Z=0.0001$) to approximate the evolution of 
metal-free stars. Since stellar wind mass loss is expected to be negligible for 
massive Pop~III stars \citep[e.g.][]{Baraffe2001}. we assume that no mass is 
lost in stellar winds. We also limit the radial expansion of
Pop~III stars. We use the evolutionary models of \cite{Marigo2001},
and impose upper limits on the radial expansion of our $Z=0.0001$ models
to approximately match those for $Z=0$ stars. The evolutionary tracks
on the Hertzsprung-Russell (H-R) diagram for our models are presented in
Figure \ref{fig.radius}. Our tracks may be directly compared with
those presented in Figure 1 of \citet{Kinugawa2014}, who were the
first to implement the results by \citet{Marigo2001} into their
population synthesis. Our tracks are slightly more luminous and extend
to slightly lower temperatures (and thus to slightly larger radii) as
compared to \citet{Kinugawa2014}.

We use the original \citet{Hurley2000} limit between Hertzsprung gap and core He
burning (CHeB); marked in Figure \ref{fig.radius}. For massive stars this distinction 
is somewhat ambiguous, as HG stars are already burning He in their cores.
This is true for both Pop~III stars and Pop~I/II stars.  
However, the initial (HG) expansion of massive stars after the main sequence (MS)
does not lead to the formation of a convective envelope. Only later evolutionary 
phases when stars become cool (during CHeB) may potentially allow the formation of
a convective envelope.  This was recently demonstrated for high- and moderate-metallicity
stars ($Z=\zsun$ and $Z=0.1\zsun$) by \citet{Pavlovskii2016}. For example,
their $80\msun$ model for $Z=0.1\zsun$ develops a convective envelope only
after exceeding \textbf{a} radius $R\approx 2000\rsun$. For smaller metallicity,
like in the case of our models, this radius should be smaller. For comparison,
our $100\msun$ model reaches the CHeB phase at a radius $R\approx 400\rsun$. 
In contrast, \citet{Kinugawa2014} includes HG phase in CHeB phase, and their 
$100\msun$ model enters the CHeB phase at $R\approx 20\rsun$ (see their Figure 1). 
This leads to much larger parameter space for entering CE phase and for the
formation of BH-BH mergers than indicated by recent detailed CE calculations 
\citep{Pavlovskii2016}.

\subsection{Black hole mass spectrum}
\label{sec.bhmass}

We employ the rapid supernova model from \citet{Fryer2012}. This model
is able to reproduce the mass gap between NSs and BHs observed in Galactic 
X-ray binaries \citep{Belczynski2012}, as well as the mass distribution of 
the three BH-BH merger events detected by LIGO if they originated from 
Pop~I/II stars \citep{Belczynski2016b}. In this model, Pop~I/II stars form 
NSs ($M_{\rm NS}=1 - 2\msun$) with energetic supernova (SN) explosions, and 
most of the star mass is ejected. Light BHs ($M_{\rm BH}\approx5 - 10\msun$; 
this is only an approximation, as the model is more complex and non-monotonic) 
are formed in low-energy SNe explosions that allow for some ejected mass
to fall back onto the central compact object. Massive BHs 
($M_{\rm BH} \gtrsim10\msun$) are formed with no (or almost no) SN explosions, 
and the entire star ends up as a BH (with some mass lost in neutrino emission). 
The specific formulas for compact object mass are presented in \citet{Fryer2012} 
and are based on star mass and CO core mass at the time of SN. Recently, we 
have updated this prescription by adding the effects of pair-instability 
supernova (PSN) and pair-instability pulsation supernova (PPSN;
\citealt{Belczynski2016c}). PSN fully disrupts the star and no remnant
is left, which limits the formation rates of double compact
objects. PPSN subjects a star to a severe mass loss just before core
collapse, and thus limits the mass of the BH formed by a
given star. Our prescription depends on the helium core mass: for
$M_{\rm He}<45\msun$ stars are not subject to PPSN nor PSN; for
$45<M_{\rm He}<65\msun$ the stars are subject to PPSN and lose all the
mass above the inner $45\msun$; for $65<M_{\rm He}<135\msun$ the stars
are subject to PSN and they do not form NS/BH, and for
$M_{\rm He}>135\msun$, despite the fact that pair-instability operates,
it is not able to overcome the gravity, and the entire star collapses to a BH
\citep[e.g.][]{Fryer2001,Heger2002,Woosley2007}.

\begin{figure}
        \hspace*{-0.5cm}
        \includegraphics[width=9.1cm]{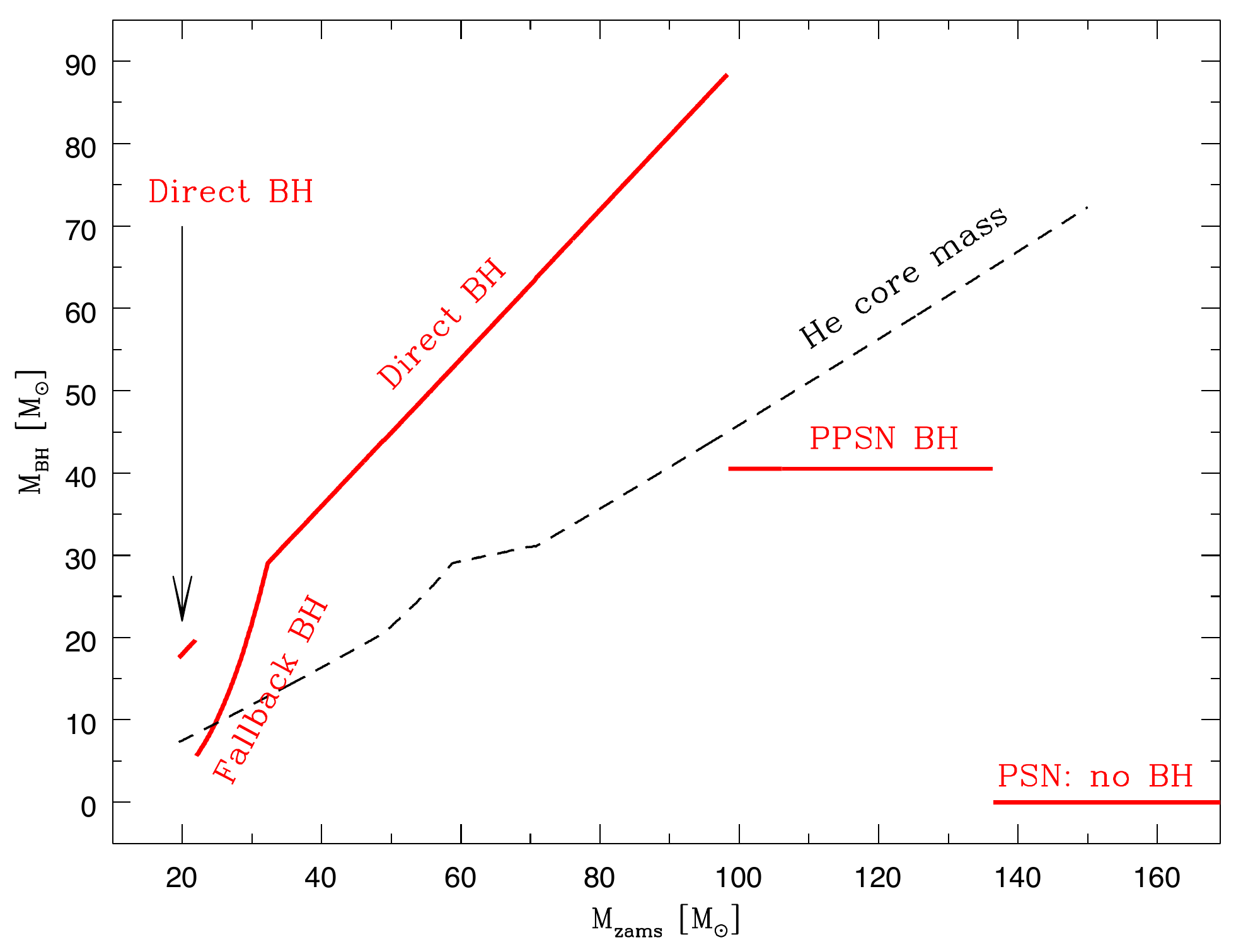} 
        \vspace*{-0.0cm}
	\caption{
		Black hole mass spectrum for our model of Pop~III stars as a
		function of stellar initial mass at Zero Age Main Sequence. We mark
		regions of fallback and direct BH formation. We also mark the region where
		stars forming BHs are subject to pair-instability pulsation supernovae 
		(PPSN), and stars that are totally disrupted by pair-instability supernovae 
		(PSN). The additional line shows the helium core mass at the time of
		core-collapse; above $M_{\rm He}>45\msun$ stars are subject to
		pair-instabilities.  
	}
	\label{fig.bhmass}
\end{figure}

We have now applied the same prescriptions to Pop~III stars. The core
collapse, and mass of a compact object, is mostly driven/set by the mass of the 
core. We use our updated Pop~III evolutionary prescriptions to get an estimate 
of Pop~III star core mass. 
In Figure \ref{fig.bhmass} we show the BH mass spectrum. Stars with initial mass range   
$19.5<M_{\rm zams}<22\msun$ form $\sim 20\msun$ BHs through the entire star 
collapsing to a BH (we assume $10\%$ mass loss through neutrino emission only). 
Stars within the mass range $22<M_{\rm zams}<33\msun$ form BHs with masses
$5$--$30\msun$ through partial fallback that is accompanied by weak SN 
explosions. Stars within the mass range $33<M_{\rm zams}<98\msun$ form the most
massive BHs: $30 - 90\msun$ through direct collapse. Stars within
the mass range $98<M_{\rm zams}<136\msun$ are subject to mass loss through
pair-instability pulsation supernovae and they form relatively low-mass
black holes with mass $\sim 40\msun$. Finally, stars above 
$M_{\rm zams}>136\msun$ are totally disrupted by pair-instability supernovae;
no BHs are formed. 

We note that we have made the strong assumption that PPSN can
remove all the mass above inner to $45\msun$ {\textit independent} of envelope
mass. This assumption is justified for Pop~I/II stars, for which the maximum 
mass of the envelope (above the inner $45\msun$) is about $\lesssim 30\msun$
\citep{Belczynski2016c} and the explosion models show that this mass may
be lifted by PPSN \citep{Woosley2015,Woosley2016}. However, for
Pop~III stars a massive envelope may be retained due to lack of
strong stellar wind mass loss. In our framework, in the PPSN regime, stars
can have mass anywhere in the range $98$--$136\msun$ and the
pair-instability pulsations may not be able to lift off all the mass
above the inner $45\msun$. A recent study has put an upper limit to what
can be retained by Pop~III stars after PPSN: the inner
$\sim 70\msun$ (S.Woosley; 2016: private communication). If this is
taken into account, the  BH mass in the PPSN regime ($98<M_{\rm zams}<136\msun$)
could be higher than what we have estimated ($\sim 40 - 70\msun$, instead of
the $40\msun$ shown in Figure~\ref{fig.bhmass}). This has no effect on our
results as during evolution leading to the BH-BH merger formation considered
in our study each massive star (BH progenitor) is stripped from its H-rich 
envelope by either stable mass transfer or by common envelope (see 
Sec.~\ref{sec.evolution}). The most massive stars in PPSN regime have helium cores 
of $\sim 65\msun$ (see Fig.~\ref{fig.bhmass}), so PPSN needs to lift off only 
outer $\sim 20\msun$.

\subsection{Pop~III Star Formation Rate}
\label{sec.pop3sfr}

In Figure~\ref{fig.sfr} we show the Pop~III star formation rate. 
We assume that the first stars form in the redshift range $z=2 - 50$. 
The Pop~III star formation rate is highly uncertain. We have adopted 
a rather optimistic model (high rate) from \cite{deSouza+11},
as \citet{Kinugawa2014} did. The Pop~III star formation rate consistent with 
the recent {\it Planck} cosmic microwave background data on optical depth to 
electron scattering in Universe is factor of $\sim 2$ lower \citep{Inayoshi2016}
than what we have adopted in this study. It means that all our merger rate and 
detection rate predictions could be factor of two lower than reported in
following sections. 

For comparison, we also show the star formation rate of Population
I/II stars.  This mode of star formation is well constrained to
redshift of $z=2$.  At higher redshifts a number of uncertainties
affect estimates. We show two estimates that clearly demonstrate the
effect of these uncertainties: the high SFR model is taken from 
\citet{Strolger2004}, and the low SFR model is taken from 
\citet{Madau2014}. We assume that Pop~I/II stars form in the redshift 
range $z=0 - 15$.

\begin{figure}
        \hspace*{-0.5cm}
        \includegraphics[width=9.1cm]{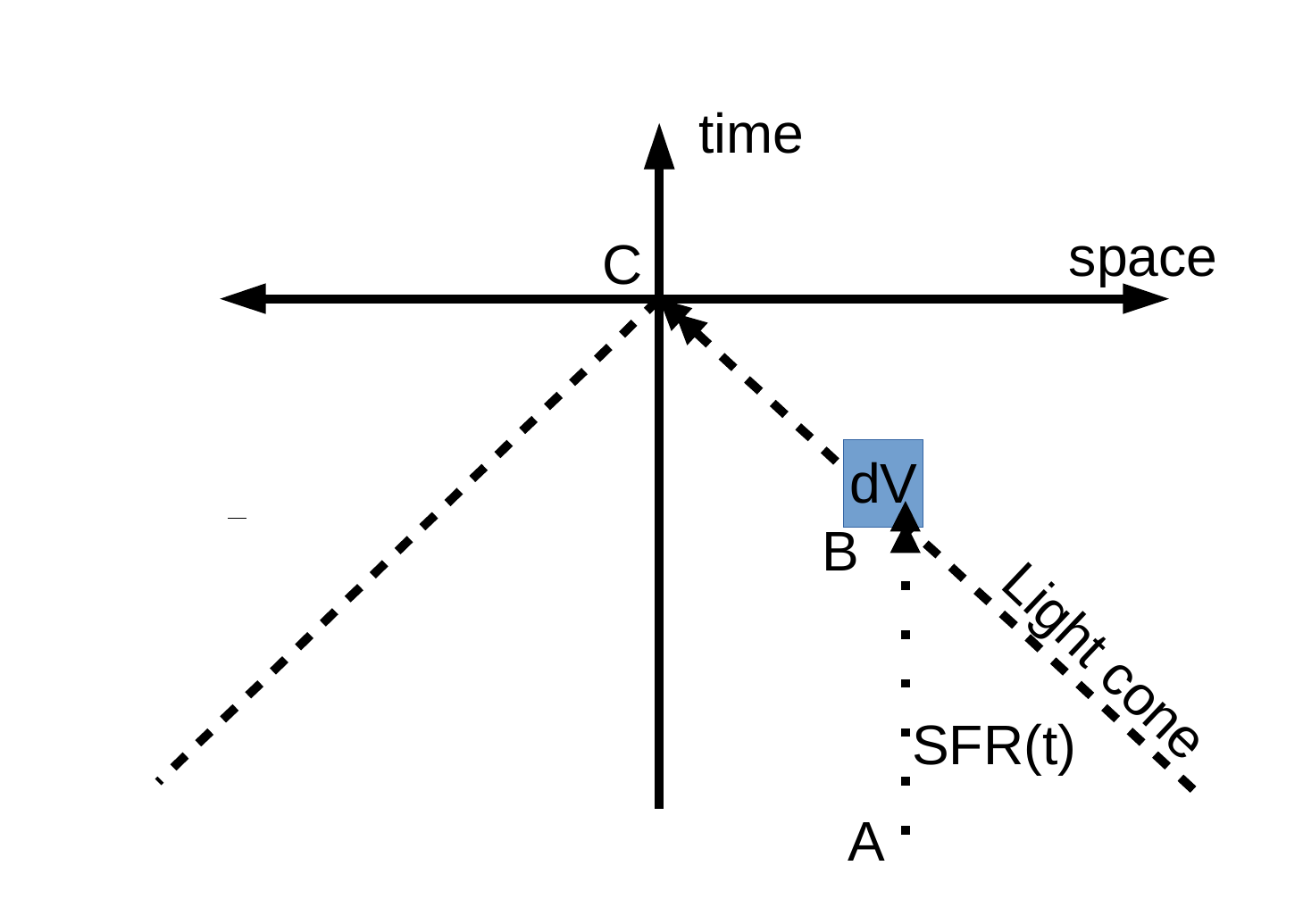} 
        \vspace*{-0.0cm}
	\caption{
        The explanation of the concept of finding the total mass of stars 
        that can lead to the formation of BH-BH mergers detectable by AdLIGO. 
        C is the present, B is the currently observable volume element, and 
        A is the distant past of this volume element (see Sec.~\ref{sec.pop3sfr} 
        for details).
	}
	\label{fig.starmass}
\end{figure}

The total stellar mass available in entire Universe for the formation of 
BH-BH, BH-NS and NS-NS mergers that are detectable by AdLIGO is given by 
the integral
\begin{equation}
M= \int_0^\infty {\rm d}z {{\rm d}V\over {\rm d}z} \int_0^{t(z)} SFR(t) ~ {\rm d}t,
\label{eq:SFRhist}
\end{equation}
where ${\rm SFR}(t)$ is the star formation rate at a given time. 
We first integrate the star formation rate to obtain the density of stars 
formed upto a given time since the Big Bang, and then integrate the 
stellar mass density with the cosmic volume to obtain the total mass.
The concept is explained in Figure~\ref{fig.starmass}. The point C corresponds 
to the present moment where the observer is located. The observer sees the 
Universe only along the light cone extending to the past denoted by dashed lines.
Thus the past of the observer contains the entire region below the dashed line. 
However only the hyper-surface denoted by the dotted line is visible to the 
observer. Nevertheless any volume ${\rm d}V$ that is seen by the observer 
contains stars or their remnants that were formed in ${\rm d}V$ during its past. 
This is denoted by the dotted line. Thus in order to find the star density we 
first integrate the star formation rate over time along the dotted line to obtain 
the density of stellar mass formed in the past of the volume element ${\rm d}V$, 
going from A to B. Then we integrate the density with the volume seen by the 
observer along the dashed line going from B to C.

The SFR formulae can be found in Belczynski et al. (2016, for Pop~I/II)and 
\citeauthor{deSouza+11} (2011, for Pop~III). Integrating eq.~\ref{eq:SFRhist}, 
we find that the total mass of Pop~III stars formed in the observable Universe 
(in the redshift range $0<z<50$) is on the order of $2.1\times 10^{18}\msun$,
whereas the total mass in Population  I/II stars (within $z<15$) is 
$8.0\times 10^{20}\msun$, so about $2.5$ orders of magnitude higher.

\begin{figure}
        \hspace*{-0.5cm}
        \includegraphics[width=9.1cm]{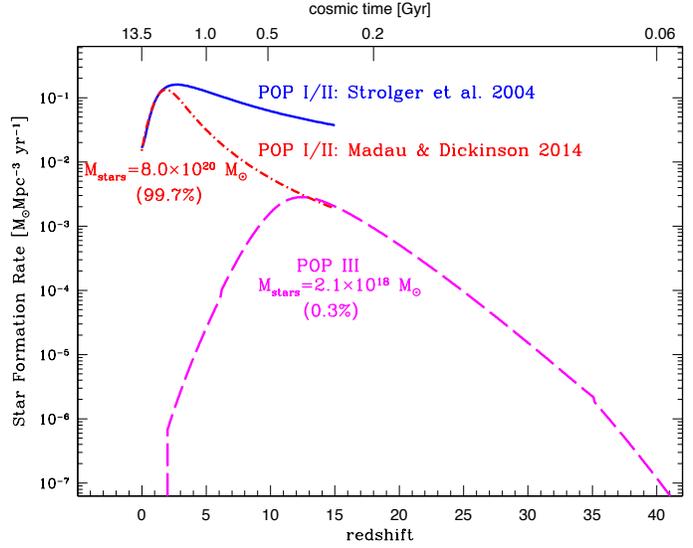} 
        \vspace*{-0.0cm}
	\caption{
		Our adopted model for the star formation rate of Pop~III stars. 
		For comparison we also show two different estimates of star formation 
                rate of Pop~I/II stars. Note that Pop~III stars form only (at most) 
                $\sim 0.3\%$ of all stars in Universe that can contribute to AdLIGO 
                detectable BH-BH mergers (see Sec.~\ref{sec.pop3sfr} for details).
	}
	\label{fig.sfr}
\end{figure}

\subsection{Initial conditions} 

\begin{figure}
        \hspace*{-0.5cm}
        \includegraphics[width=9.1cm]{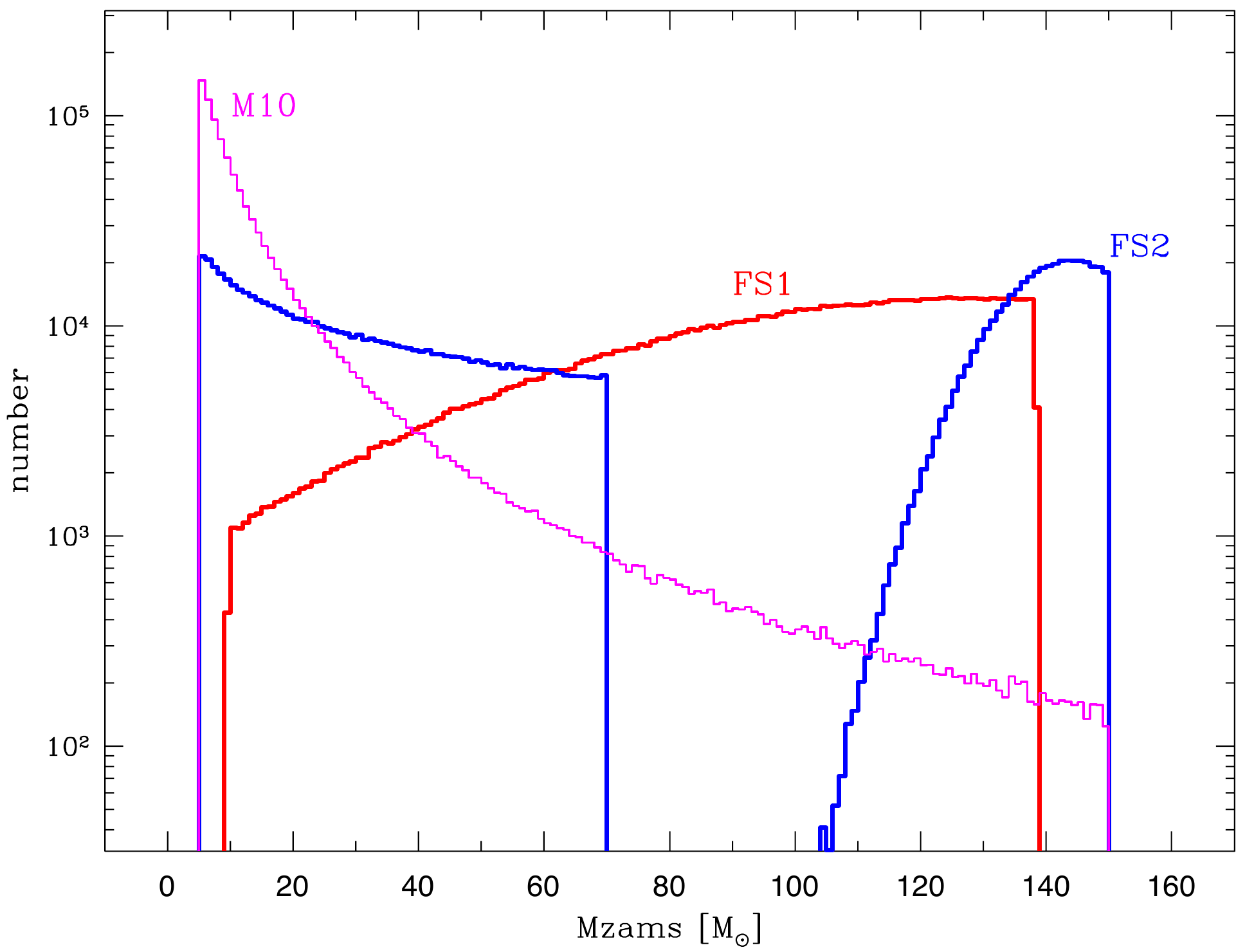} 
        \vspace*{-0.0cm}
                \caption{ Initial mass function of primary stars in
                  Pop~III binaries. Our two basic models of Pop~III
                  binary formation (FS1 and FS2) are shown.  For
                  comparison we show the initial mass function that is
                  typical for massive Pop~I/II stars (model M10:
                  $\propto M_{\rm zams}^{-2.3}$).}
	\label{fig.imf}
\end{figure}

\begin{figure}
        \hspace*{-0.5cm}
        \includegraphics[width=9.1cm]{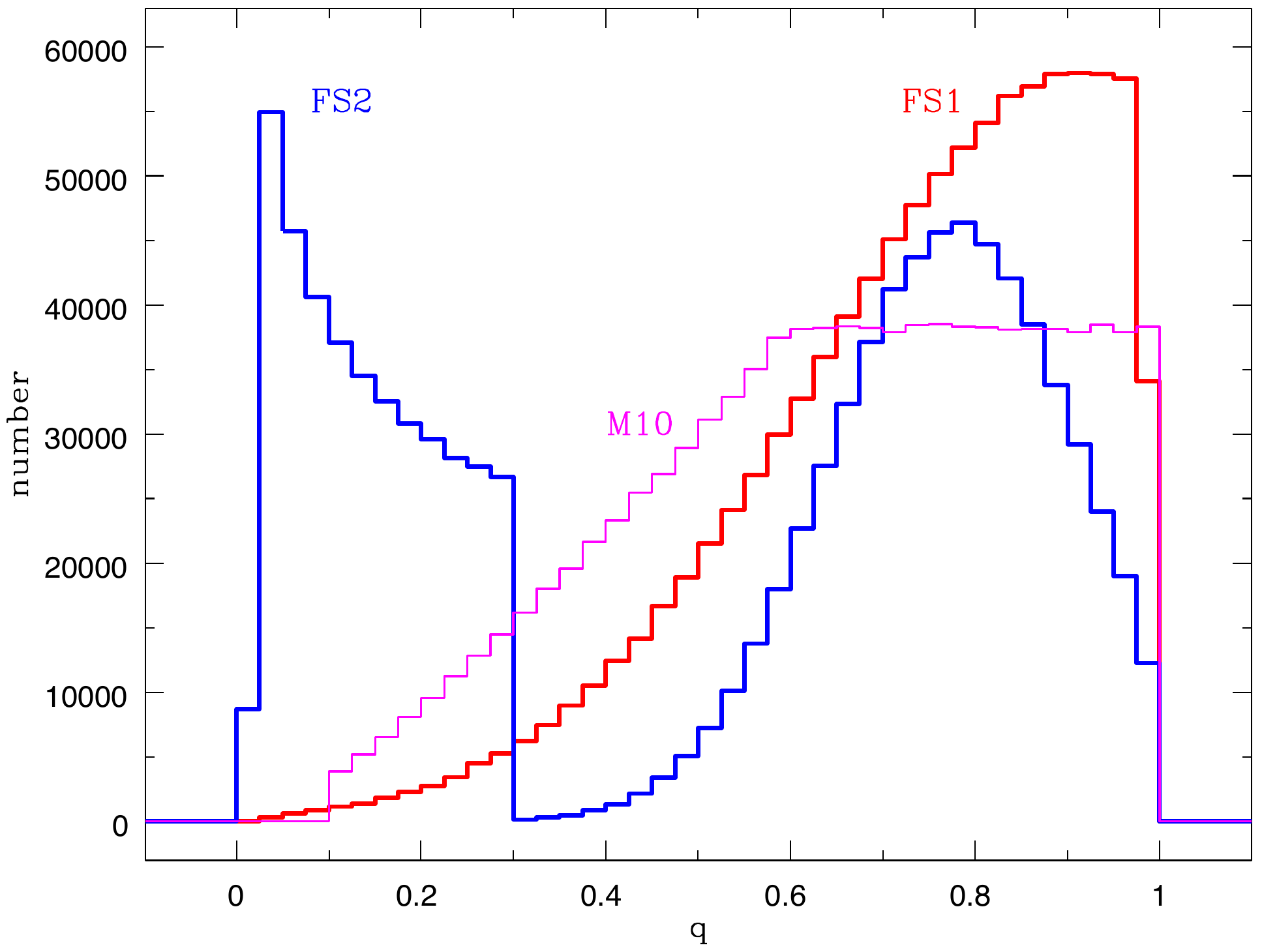} 
        \vspace*{-0.0cm}
	\caption{
		Initial mass ratio (secondary---to---primary) distribution for Pop~III 
		binaries. Our two basic models of Pop~III binary formation (FS1 and FS2) 
		are shown. For comparison we show the initial mass ratio typical for
		massive Pop~I/II stars (model M10): underlying mass ratio distribution 
		is flat (see Tab.~\ref{tab.init}). However, since we only evolve stars 
                above certain limit on primary mass ($>5\msun$) and secondary mass 
                ($>3\msun$) it causes a depression of low mass-ratio systems in our 
                simulated binaries.  
	}
	\label{fig.q0}
\end{figure}

\begin{figure}
        \hspace*{-0.5cm}
        \includegraphics[width=9.1cm]{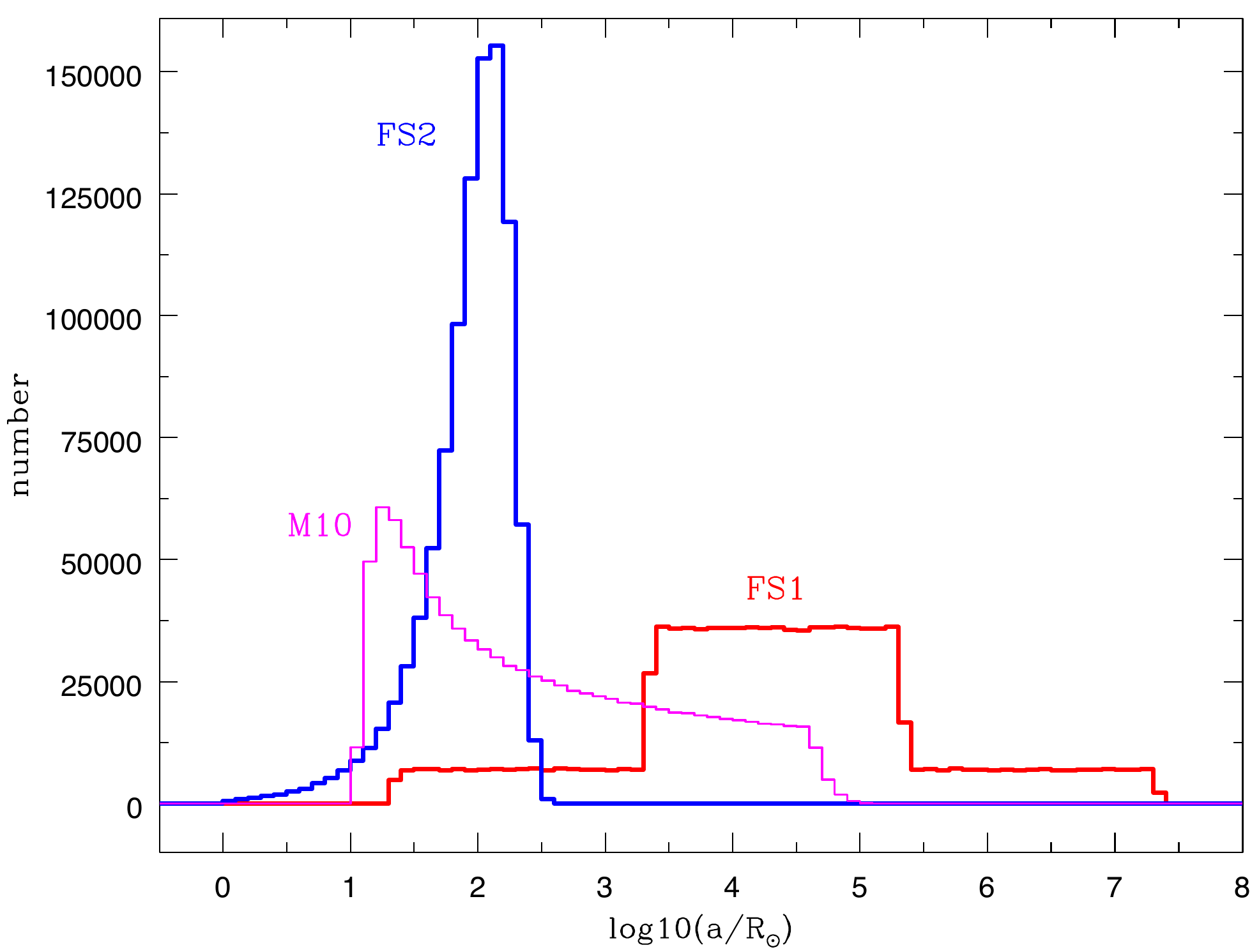} 
        \vspace*{-0.0cm}
	\caption{
		Initial orbital separation distribution for Pop~III binaries. Our 
		two basic models of Pop~III binary formation (FS1 and FS2) are shown. 
		For comparison we show the initial orbital separations typical for          
		massive Pop~I/II stars (model M10).
	}
	\label{fig.a0}
\end{figure}

\begin{figure}
        \hspace*{-0.5cm}
        \includegraphics[width=9.1cm]{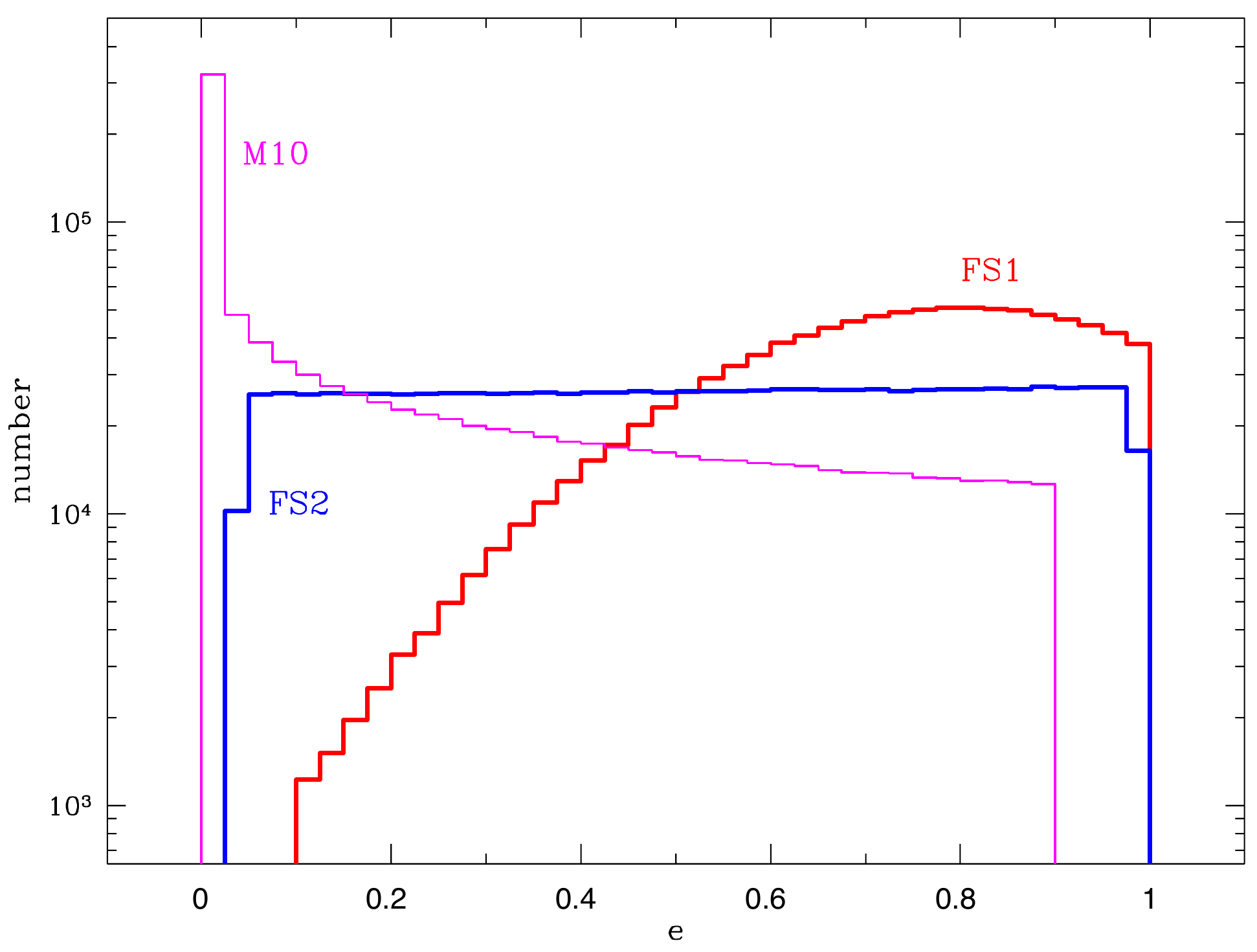} 
        \vspace*{-0.0cm}
	\caption{
		Initial orbital eccentricity distribution for Pop~III binaries. Our 
		two basic models of Pop~III binary formation (FS1 and FS2) are shown. 
		For comparison we show the initial orbital eccentricities typical for 
		massive Pop~I/II stars (model M10).
	}
	\label{fig.e0}
\end{figure}

In Table~\ref{tab.init} we provide the details of our adopted initial binary
distributions that are used for the evolution of Pop~III binaries. The
origin of these distributions is presented in Section~\ref{sec: InitialProperties}.
In Figures~\ref{fig.imf},~\ref{fig.q0},~\ref{fig.a0} and \ref{fig.e0}, we show
these initial distributions.

\begin{table*}
	\caption{Initial Conditions for Evolution of Pop~III Binaries}
	\label{tab.init}
		\centering
		\begin{threeparttable}
		\begin{tabular}{c| c c c c}
		\hline\hline
		model & IMF\tnote{a} & $q$\tnote{b} & $a$ or $P_{\rm orb}$\tnote{c}& $e$\tnote{d}\\
		\hline
		& Gaussian                           & Gaussian                  & flat in $\log(a)$                               & Gaussian \\ 
		FS1& $\sigma=52.2\msun$, $M_0=128\msun$ & $\sigma=0.29$, $q_0=0.92$ & $72\%$ in range1: $2000$--$2\times10^5\rsun$    & $\sigma=0.25$, $e_0=0.8$ \\  
		& range: $9.6$--$138\msun$           & range: $0.03$--$0.99$     & $28\%$ outside range1: $20$--$2\times10^8\rsun$ & range: $0.10-1.0$ \\
		
		\hline 
		& power-law + Gaussian               & power-law + Gaussian                & Gaussian                              & linear \\ 
		& $50\%$ in range1: $3$--$70\msun$   & range1: $0.002$-$0.3$ ($M>70\msun$) & $\sigma=71.6\rsun$, $a_0=90.1\rsun$   & slope $=0.08$ \\  
		FS2& $\alpha=-0.55$                     & $\alpha=-0.35$                      & range: $1.1$--$1075\rsun$             & range: $0.04$--$0.99$ \\
		& $50\%$ in range2: $70$--$181\msun$ & range2: $0.1$--$1.0$ ($M<70\msun$)  &                                       & \\
		& $\sigma=11.0\msun$, $M_0=144\msun$ & $\sigma=0.14$, $M_0=0.78$           &                                       & \\
		\hline
		& power-law                          & flat                                & power-law in $\log(P_{\rm orb}/day)$  & power-law \\ 
		M10& $\alpha=-2.3$                      &                                     & $\alpha=-0.5$                         & $\alpha=-0.42$ \\
		& range: $0.08$--$150\msun$          & range: $0.1$--$1$                   & range: $0.15$--$5.5$                  & range: $0.0$--$0.9$ \\
		
		\hline
		\hline
		
	\end{tabular}
	\begin{tablenotes}
		\item[a] Initial mass function: shape and range
		\item[b] Mass ratio distribution: shape and range
		\item[c] Orbital separation or period distribution: shape and range
	    \item[d] Eccentricity distribution: shape and range
		\end{tablenotes}
\end{threeparttable}
\end{table*}

\section{Evolutionary channels}
\label{sec.evolution}

\subsection{Model FS1: moderate orbital separations}
\label{sec.pp1}

In this model majority of BH-BH mergers are formed along one specific
evolutionary channel (see Table~\ref{tab.evol}). Since most of the initial
binaries have similar mass components (see mass ratio distribution in 
Fig.~\ref{fig.q0}) and since all stars follow similar evolution (specific 
to zero metallicity) we find that BH-BH mergers form predominantly along
just one evolutionary sequence. A typical example of the evolution is given 
in the following.  

Evolution begins with two massive, very similar, stars $M_{\rm 1}=69.6\msun$ and 
$M_{\rm 2}=65.6\msun$ on a moderately wide ($a=9955\rsun$) and rather eccentric 
orbit ($e=0.73$). Not only initial mass ratio peaks at high values $q \sim 0.8 - 1$
(Fig.~\ref{fig.q0}), but also systems with very low mass ratios tend to evolve 
through two CE phases and they are very likely to merge  during one of the
CE event, barring the formation of BH-BH merger. 

The first Roche-lobe overflow (RLOF) phase starts while both stars are already evolved 
core helium burning stars (CHeBs). The onset of RLOF begins when the
stars meet at periastron ($d_{\rm per}=a(1-e)=2690\rsun$). The orbit is not
circularized by tidal forces due to relatively weak tidal efficiency noted in 
massive stars \citep{Claret2007}, and due to the very short time of $0.2$ Myr 
since the primary star left MS ($R_{\rm 1}=20\rsun$) until it has reached RLOF 
during CHeB ($R_{\rm 1}=1073\rsun$). At the onset of RLOF we instantaneously
circularize the orbit at periastron ($a=2690\rsun$, $e=0$) assuming that
effective dissipation will take place at periastron passages. The RLOF
develops into stable but non-conservative mass transfer (half of the mass lost by
the donor is ejected from the system). Very quickly, the mass ratio is reversed, and
now the donor primary is the least massive star. The mass lost from the system
carries off specific angular momentum ($j_{\rm loss}=1.0$ defined in
\citealt{Podsiadlowski1992}) and the orbit expands. At the time when the primary
is depleted to $M_{\rm 1}=43.7\msun$ the orbit has expanded to $a=3436\rsun$. At
the same time the secondary star, which is also on CHeB, has gained mass to 
$M_{\rm 2}=78.6\msun$ and its radius has reached its Roche lobe ($R_{\rm 2}=1534\rsun$). 
At this point the system enters a double common envelope (CE) phase with two
helium cores inspiraling within the two H-rich envelopes of both stars. 

We perform CE with $\alpha=100\%$ efficient energy transfer from the orbit into 
ejection of the primary and secondary envelopes. We use a physical estimate of 
the envelope binding energy with $\lambda=0.07$ and $0.05$ for the primary and the
secondary, respectively \citep{Xu2010,Dominik2012}, i.e. the envelopes
are $14 - 20$ times more bound to their cores (and harder to eject)
than typically assumed in Pop~III population synthesis studies
(e.g., \citealt{Kinugawa2014} uses $\lambda=1.0$). After the ejection
of the massive double envelope ($64.0\msun$) the orbital separation
decreases to $a=6.7\rsun$, and it hosts two massive naked helium cores
($M_{\rm 1}=30.2\msun$, $M_{\rm 2}=28.0\msun$).

After CE, both helium cores evolve toward core-collapse. At $t=4.3$ Myr
after binary formation, the primary undergoes core-collapse. Its 
CO core mass is $M_{\rm 1,co}=24.0\msun$ and the BH forms through direct
collapse of a star to a BH with no mass ejection and no natal kick \citep{Fryer2012}. 
We only assume $10\%$ mass loss in neutrino emission, which induces a small
orbital widening and small eccentricity of the system ($a=7.1\rsun$, $e=0.05$).
At $t=4.4$ Myr since binary formation, the secondary undergoes
core-collapse. Its CO core mass is $M_{\rm 1,co}=22.1\msun$ and the BH forms 
with no mass ejection and no natal  kick \citep{Fryer2012}.
We again assume $10\%$ mass loss in neutrino emission, 
which induces further orbital widening and increases the eccentricity of the system 
($a=7.5\rsun$, $e=0.08$). The two massive black holes $M_{\rm 1,bh}=27.1\msun$ and 
$M_{\rm 2,bh}=25.2\msun$ have formed on a very close orbit with a total delay
time of $17$ Myr  (evolutionary time of $4.3$ Myr and merger time of $12.7$ Myr).

\subsection{Model FS2: small orbital separations}
\label{sec.pp2}

In this model the formation of BH-BH mergers is almost totally suppressed.
The only evolutionary channel that allows the formation of BH-BH mergers 
(described below) is most likely an artifact that emerged due to limitations 
of our evolutionary model. At best, even if this unlikely scenario works, due 
to very low BH-BH merger formation rate and short merger times, the BH-BH 
merger rate within Advanced LIGO horizon is zero. 

We describe only one particular system that has formed BH-BH merger in this
model framework. The other systems have very similar initial properties and 
the same evolutionary sequence. 
Evolution begins with two massive stars $M_1=67.7\msun$ and $M_2=52.1\msun$ 
on a relatively close ($a=36.3\rsun$) and almost circular orbit ($e=0.12$). 
The first RLOF starts while both stars are still on MS and develops into stable 
but non-conservative mass transfer. The donor, the more massive primary, keeps 
losing mass as it evolves off MS, through Hertzsprung gap (HG) and becomes a CHeB. 
At the end of the mass transfer phase, the primary has lost most of its mass
($M_1=12.5\msun$), half of which was accreted onto and rejuvenated the
secondary star ($M_2=79.7\msun$) and half is lost from the binary. The mass
transfer leads to the orbital expansion and circularization ($a=208\rsun$,
$e=0$).

The primary, which lost most of its H-rich envelope during CHeB, becomes a naked
helium (He) star and quickly evolves toward core-collapse. At $t=4.8$ Myr,
the primary explodes in Type Ib/c supernova. Its CO core mass is 
$M_{\rm 1,co}=9.3\msun$ and the BH that forms is subject to mass ejection
and natal kick \citep{Fryer2012}. The combined effects of mass ejection
and natal kick place a newly formed black hole ($M_{\rm 1,bh}=8.7\msun$) on
an extremely wide and eccentric orbit ($a=1.6 \times 10^6 \rsun$, $e=0.999$).
At closest approach (periastron) the orbital separation is $a_{\rm per}=1600\rsun$, 
with primary and secondary Roche lobe radii of $R_{\rm 1,rl}=345\rsun$ and 
$R_{\rm 2,rl}=927\rsun$, respectively. 

The secondary evolves off MS, goes through HG and evolves along CHeB until
its size becomes larger than its periastron Roche lobe radius at time $t=5.6$ 
Myr. We {\textit assume} that this eccentric system circularizes at one of the
periastron passages where the dissipation of tidal energy is expected to be
efficient (primary overfilling its instantaneous Roche lobe). We reset the 
system on the new orbit 
($a=a_{\rm per}=1600\rsun$  and $e=0$) at the first periastron passage in
which the secondary overfills its Roche lobe. Then we check for the stability 
of ensuing RLOF. Due to the large mass ratio ($M_{\rm 1,bh}=8.7\msun$ and 
$M_2=79.7\msun$) and the fact that the donor is a CHeB star with convective
envelope, such system is expected to evolve through common envelope phase. 
We perform CE with $\alpha=100\%$ efficient energy transfer from the orbit 
into ejection of the secondary envelope. We estimate the 
envelope  binding energy using $\lambda=0.2$ \citep{Xu2010}, which means that
the envelope is $5$ times more bound to its core (and harder to eject) than
typically assumed in Pop~III population synthesis studies.
 After the ejection of the massive
envelope ($45.2\msun$), the orbital separation decreases to $a=8.3\rsun$, and
it hosts a black hole with increased mass ($M_{\rm 1,bh}=9.3\msun$; 
accretion in CE) and compact massive helium core of the secondary 
($M_2=34.5\msun$). 

The secondary subsequently evolves through nuclear burning, and at
$t=5.8$ Myr with its massive CO core ($M_{\rm 1,co}=27.8\msun$) it
collapses directly to a black hole ($M_{\rm 2,bh}=31.0\msun$). In our
standard model we assume that natal kicks are associated with mass
ejection/supernova and this massive star does not explode, so it is
not subject to a natal kick.  We assume that $10\%$ of the 
mass of the secondary is lost in neutrinos, slightly increasing the
orbital separation ($a=9.1\rsun$) and inducing a small eccentricity
($e=0.09$). The two BHs merge at $t_{\rm mer}=86$ Myr. The time
between star formation and binary merger is very short (evolutionary
time of 5.8 Myr and merger time of 86 Myr): $t_{\rm del}=92$
Myr. Because Pop~III star formation ends at $z=2$, we do not expect
any BH-BH mergers within the detection range of Advanced LIGO.

Apparently, this peculiar scenario is the only way for Pop~III stars 
with these specific initial conditions (relatively small separations: 
$\lesssim 300\rsun$ see Fig.~\ref{fig.a0}) and the adopted evolutionary 
model to bring a black hole and a massive secondary to the orbit of about 
a few thousands solar radii at the onset of CE. If the initial orbit is
larger, then CE is not able to reduce the separation
below about $50 \rsun$ and the BH-BH system that forms has
merger time larger than the Hubble time. If the pre-CE orbit is
smaller than about $1000\rsun$, then CE leads to merger of a BH with
He core of the secondary.  The most uncertain part of this
evolutionary scenario is the pre-CE circularization from almost fully
elliptical orbit ($e=0.999$) to fully circular orbit ($e=0.0$). Even
if it works as we have assumed, the resulting merger rate of BH-BH
systems is very small at hight redshifts (peaking at $z=10$ at
$0.01\gpy$), and it is zero at redshifts reachable by Advanced LIGO
($z<2$).

\subsection{Model M10: Pop~I/II stars}

Model M10 is our reference model for the evolution of Pop~I/II stars.
In this model we calculate the evolution of isolated binary stars
(without dynamical interactions) and for slow and mildly rotating
stars (homogeneous evolution for rapidly spinning stars is not
included). This model is a modification of model M1, first introduced
in \citet{Belczynski2016b}, and later updated to include the effects
of pair-instability supernovae and pair-instability pulsation
supernovae as discussed by \citet{Belczynski2016c}. The pair-instability
supernovae limit the BH formation from the most massive stars, while the 
pair-instability pulsation supernovae impose a severe upper limit on the 
maximum BH mass ($\lesssim 50 \msun$) for Population I/II stars.

The typical evolution of binaries that form BH-BH mergers starts with two
massive stars and wide orbits, and proceeds in the order of stable mass
transfer, BH formation, CE evolution and second BH formation (e.g., see Fig.~1
of \citealt{Belczynski2016b}). A summary of this sequence, along with the 
Pop~III formation channels, is given in Table~\ref{tab.evol}.

\begin{table}
\caption{Major formation channels of BH-BH mergers}
\label{tab.evol}
\centering
\begin{threeparttable}
\begin{tabular}{c| l}
\hline \hline
Model & Evolutionary sequence\tnote{a} \\
\hline 
FS1 &  MT1(4-4)\ \ \ CE12(4-4;7-7)\ \ \ BH1\ \ \ BH2 \\ 
\hline
FS2 &  MT1(1/2/4-1)\ \ \ BH1\ \ \ CE2(14-4;14-7)\ \ \ BH2 \\
\hline
M10 &  MT1(2/4-1)\ \ \ BH1\ \ \ CE2(14-4;14-7)\ \ \ BH2 \\
\hline
    & MT1(4-4)\ \ \ CE12(4-4;7-7)\ \ \ BH1 BH2 \\
KK1 & MT1(4-4)\ \ \ CE2(7-4;7-7)\ \ \ BH1\ \ \ BH2 \\
    & MT1(2/4-1)\ \ \ BH1\ \ \ CE2(14-4;14-7)\ \ \ BH2 \\
\hline 
    & MT1(1/2/4-1)\ \ \ BH1\ \ \ CE2(14-2/4;14-7)\ \ \ BH2 \\
KK2 & CE1(4-1;7-1)\ \ \ BH1\ \ \ CE2(14-4;14-7)\ \ \ BH2 \\
    & CE1(4-1;7-1)\ \ \ CE2(7-2;7-7)\ \ \ BH1\ \ \ BH2 \\    
\hline \hline
\end{tabular}
\begin{tablenotes}
\item[a]
MT -- stable mass transfer, CE -- common envelope, BH -- core collapse and black 
hole formation initiated either by the primary star ($1$; initially more massive 
binary component) or the secondary star ($2$) or by both stars together ($12$). 
The evolutionary stage of each of the interacting components is marked in parentheses: 
main sequence star ($1$), Hertzsprung gap star ($2$), core He-burning star ($4$), 
helium star ($7$), and black hole ($14$), with the primary star listed first. For 
common envelope evolution, the first pair of numbers lists the components' 
evolutionary stage before CE, while the second pair indicates the evolutionary 
stage after CE.
\end{tablenotes}
\end{threeparttable}
\end{table}

\section{Results} 

\begin{table}
	\caption{Results of Population Synthesis Calculations}
	\label{tab.eff}
	\centering
	\begin{threeparttable}

	\begin{tabular}{c| c c c c c}
		\hline\hline

model & $X_{\rm BHBH}$\tnote{a} & $M_{\rm sim}$\tnote{b} & 
$N_{\rm BHBH}$\tnote{c} & $N_{\rm BHNS}$\tnote{c} & 
$N_{\rm NSNS}$\tnote{c}\\
&&   $10^9 M_\odot$ &&&\\
\hline
FS1     & $9.5 \times 10^{-5}$ & $3.5$ &   $332,003$ &    $56$ &     $0$ \\
FS2     & $3.8 \times 10^{-9}$ & $1.6$ &         $6$ &     $0$ &     $0$ \\
\hline
M10\tnote{d}  &&&&&\\
0.0002  & $3.0 \times 10^{-5}$ & $1.4$ &    $41,575$ &   $750$ & $2,580$ \\
0.002   & $1.1 \times 10^{-5}$ & $1.4$ &    $16,080$ & $2,325$ & $2,275$ \\
0.02    & $1.9 \times 10^{-7}$ & $1.4$ &       $270$ &    $65$ & $7,630$ \\

\hline
KK1     & $3.5 \times 10^{-5}$ & $2.0$ &    $70,353$ &    $87$ &     $0$ \\
KK2     & $5.8 \times 10^{-4}$ & $2.0$ & $1,162,155$ & $1,224$ &     $0$ \\		
		\hline \hline
		
	\end{tabular}

	\begin{tablenotes}
		\item[a] BH-BH merger formation efficiency per unit of star
		forming mass: $X_{\rm BHBH}=N_{\rm BHBH}/M_{\rm sim}$
		\item[b] Total mass of stars across entire IMF (single, binaries, 
		triples) corresponding to a given simulation.
		\item[c] Number of BH-BH, BH-NS and NS-NS binaries formed with delay 
		time below Hubble time
		\item[d] Model M10 for Pop~I/II stars is obtained with
		non-trivial combination of $32$ different metallicity models 
		(in range $Z=0.03 - 0.0001$). Here we show three representative metallicity
		models from M10: $Z=0.0002,002,02$.
		\end{tablenotes}
\end{threeparttable}
\end{table}

\subsection{BH-BH merger formation efficiency}

In Table~\ref{tab.eff} we list the number of BH-BH, BH-NS, NS-NS mergers formed
in each simulated stellar population model. The mass ($M_{\rm sim}$) includes 
mass in single stars and binary stars (and in triples for models FS1 and FS2) in 
the entire IMF ranges listed in Table~\ref{tab.init}. We also translate these 
numbers into formation efficiency (per unit mass) for BH-BH mergers. 

For Pop~I/II stars (model M10), the formation efficiency of BH-BH mergers 
increases with decreasing metallicity due to the higher BH masses and easier CE 
development/survival at low metallicity \citep{Belczynski2010a}. For very
low metallicity ($Z=0.0002$) this efficiency is high: at the level of $3.0 \times
10^{-5} \msun^{-1}$. 

For Pop~III stars in model FS1, the efficiency of BH-BH mergers is even
higher: $9.5 \times 10^{-5} \msun^{-1}$. This comes from the fact that the IMF in
model FS1 favors BH formation: the number of massive stars increases with initial
star mass. This is very specific for Pop~III star formation and
provides a major boost to BH-BH merger formation. 
For comparison, for Pop~I/II stars (M10) the number of massive stars 
decreases with initial star mass (see Fig.~\ref{fig.imf}). 

For Pop~III stars in model FS2 the efficiency of BH-BH mergers
is very low: $3.8 \times 10^{-9} \msun^{-1}$. In this model, initial
orbital separations do not reach the typical large
separations that are required to form BH-BH mergers
($a\gtrsim 1000\rsun$; see \citealt{deMink2015}). The large
separations are required for slow to moderately rotating stars that do
not undergo homogeneous evolution. As explained in
Section~\ref{sec.pp2}, the formation of BH-BH mergers follows only from
a supernova injection of the first BH on a very wide (and eccentric)
orbit, that after the circularization and CE evolution, produces a
BH-BH merger. This is a very unusual process with extremely low
formation efficiency.

\subsection{BH-BH merger mass}

\begin{figure}
        \hspace*{-0.5cm}
	\includegraphics[width=9.1cm]{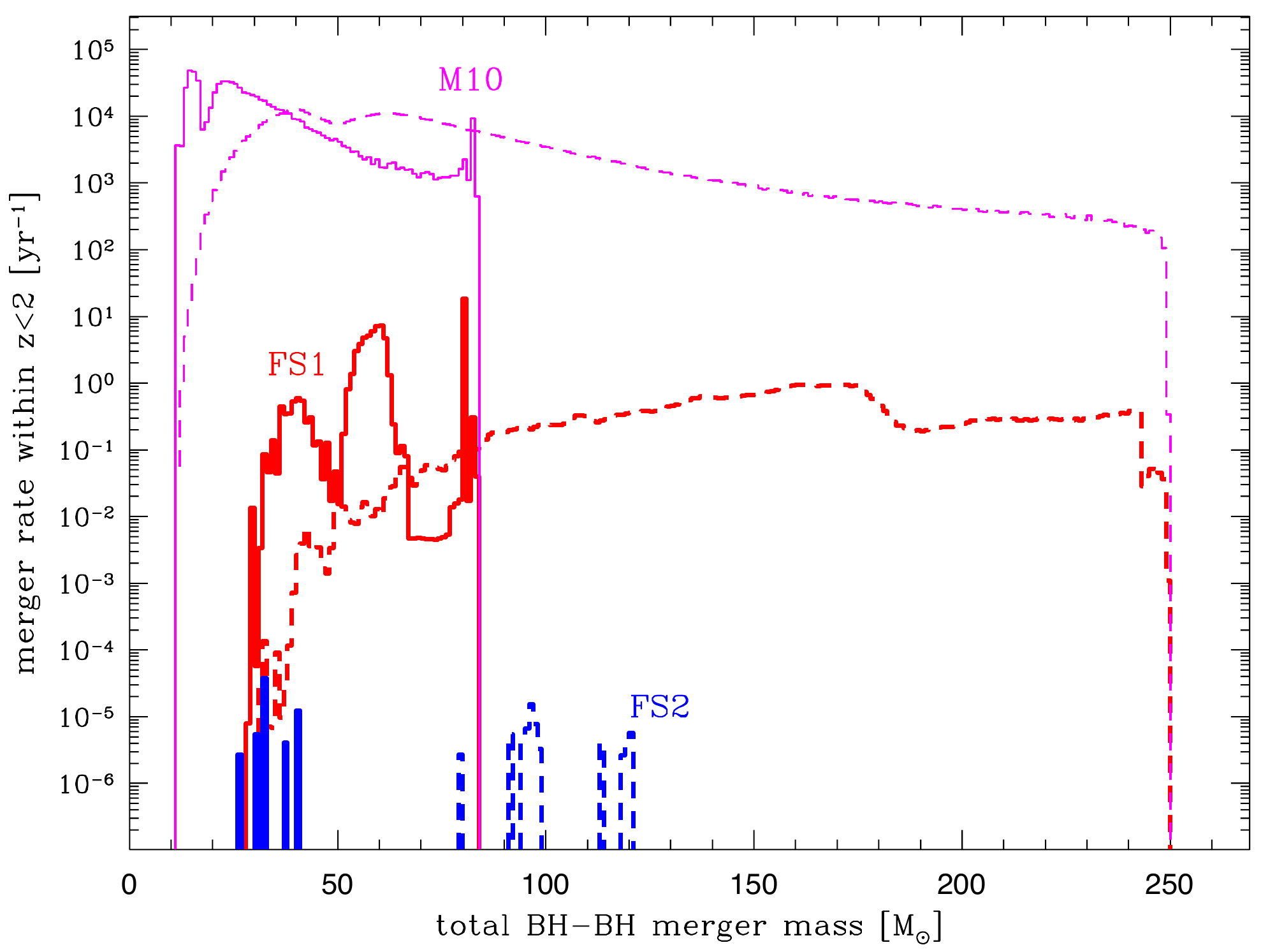}
        \vspace*{-0.0cm}
	\caption{
		Total intrinsic (solid lines) and redshifted (dashed lines) BH-BH merger mass 
		for BH-BH  mergers that take place within redshift of 2 ($z<2$). Note
		significantly higher average mass for Pop~III BH-BH mergers: 
		$M_{\rm tot}=63.4\msun$ ($M_{\rm tot,z}=162\msun$) for model FS1, as compared 
		with the Pop~I/II mergers: $M_{\rm tot}=29.7\msun$ ($M_{\rm tot,z}=73.7\msun$) 
		for model M10. For model FS2 the corresponding values are: $M_{\rm tot}=34.1\msun$
		($M_{\rm tot,z}=101\msun$). 
	}
	\label{fig.totmass}
\end{figure}

In Figure~\ref{fig.totmass} we show the distribution of total BH-BH merger mass. 
We include only systems that merge within reach of advanced LIGO at full
design sensitivity: $z<2$. We show total intrinsic mass 
($M_{\rm tot}=M_{\rm BH1} + M_{\rm BH2}$) and total redshifted mass 
($M_{\rm tot,z}=(1+z)M_{\rm tot}$) that is observed in gravitational wave 
detectors like LIGO or VIRGO. 

The intrinsic mass of Pop~III BH-BH mergers in model FS1 is found in the 
range $M_{\rm tot} \approx 30 - 80\msun$. The average total intrinsic mass 
is $63.4\msun$. 
Stars that form BH-BH mergers evolve through stable mass transfer and CE 
(see Tab.~\ref{tab.evol}). These events remove H-rich envelopes from massive 
stars. Therefore, even in case of direct BH formation, the BH mass is limited 
by the mass of the He core. This, along with pair-instability pulsation supernova 
mass loss, sets a maximum BH mass in BH-BH merger of $M_{\rm BH,max}\approx45\msun$, 
while single Pop~III stars that retain an H-rich envelope may form much more 
massive BHs $M_{\rm BH,max}=90\msun$ (see Fig.~\ref{fig.bhmass}). 

The total redshifted BH-BH merger mass in model FS1 is found in a broad 
range $M_{\rm tot,z} \approx 30 - 250\msun$. This is the result of significant 
redshifting for BH-BH mergers as the BH-BH merger rate density increases with
redshift in the entire considered redshift range ($z=0 - 2$; see Sec.~\ref{sec.rate}). 

In comparison, Pop~I/II BH-BH mergers (M10) are found in a broader mass 
range $M_{\rm tot} \approx 10 - 80\msun$ and with significantly lower average 
total intrinsic mass of $29.7\msun$. This follows from the combination of two things. 
First, Pop~I/II stars are subject to significant wind mass loss, which reduces the 
average BH mass in comparison with Pop~III evolution. Second, the significant radial 
expansion experienced by Pop~I/II stars leads to early interactions (envelope 
removal) and thus smaller masses for the He and CO cores compared with Pop~III stars, 
which are more compact.

\begin{figure}
        \hspace*{-0.5cm}
        \includegraphics[width=9.1cm]{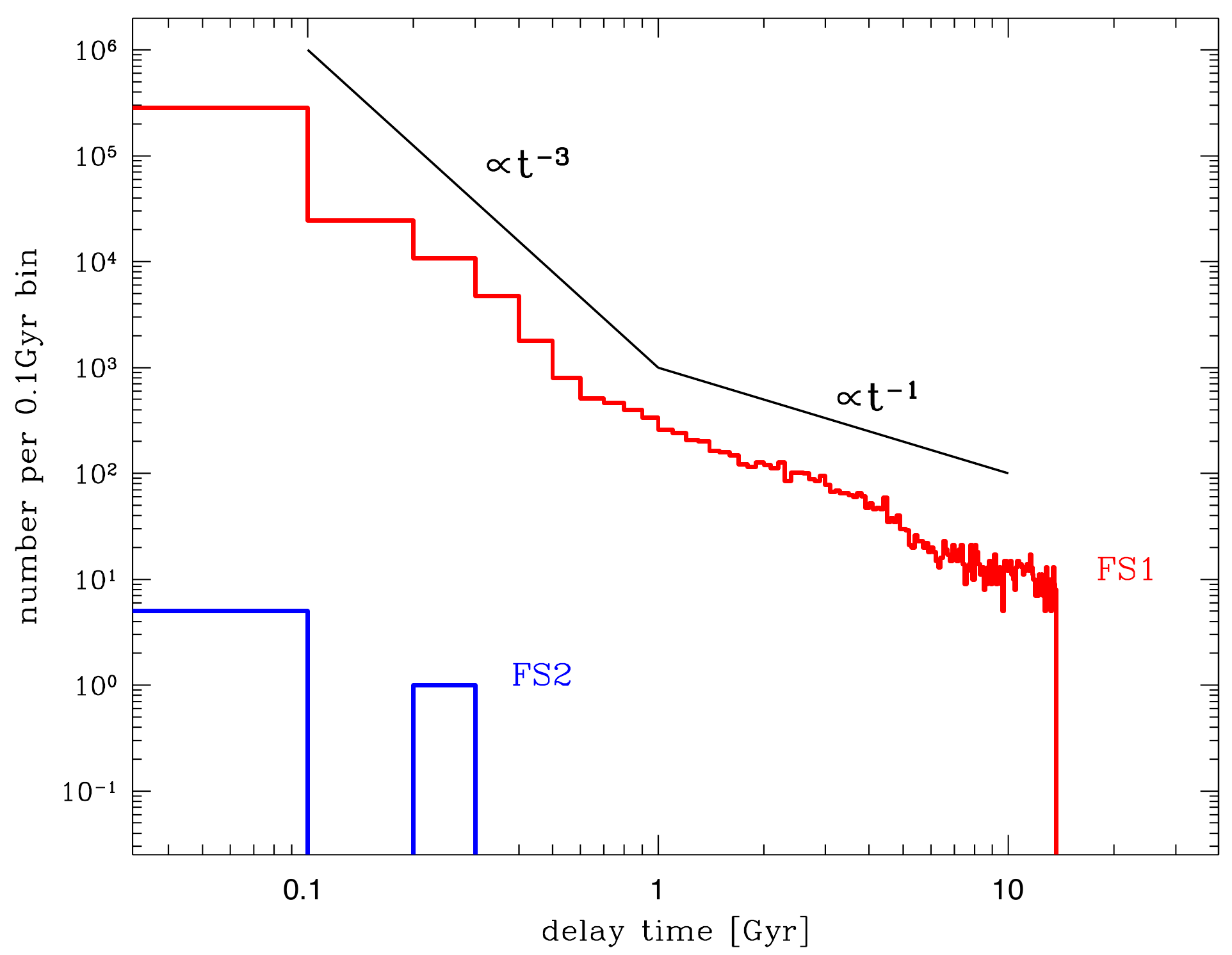} 
        \vspace*{-0.0cm}
	\caption{
		Delay times for BH-BH mergers for our Pop~III models. It 
		can be seen clearly that the delay times (from star formation to the merger) 
		are predominantly short; the average delay time is $117$ Myr for model FS1
		and $93$ Myr for model FS2. Since Pop~III star formation ended 
		around redshift of $z=2$ ($\sim 10$ Gyr ago), only the very tail of
		the distribution ($t_{\rm delay}>10$ Gyr) will contribute to the local Universe 
		($z\approx0$) merger rate. 
	}
	\label{fig.del}
\end{figure}

\begin{figure}
        \hspace*{-0.5cm}
        \includegraphics[width=9.1cm]{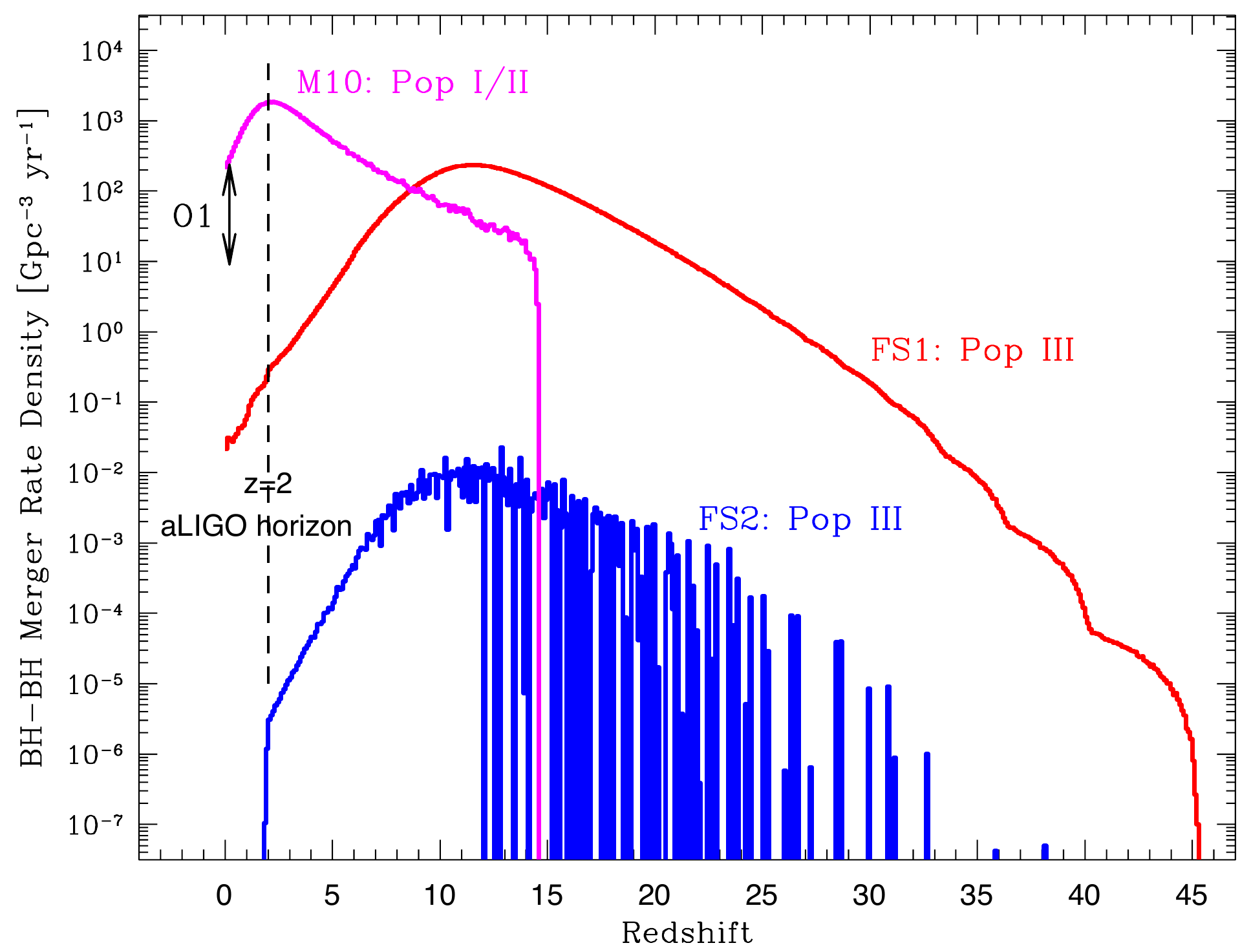} 
        \vspace*{-0.0cm}
	\caption{
	Merger rate density for BH-BH binaries as a function of redshift. 
        Two models for Pop III BH-BH mergers are shown: FS1 and FS2. For
        comparison we also show merger rate density for Pop I/II stars:
        model M10. The local BH-BH merger rate density measured by advanced 
        LIGO during O1 observations is marked. We also mark advanced LIGO 
        horizon for its full (design) sensitivity.          
	}
	\label{fig.rates}
\end{figure}

\begin{figure}
	\includegraphics[width=\columnwidth]{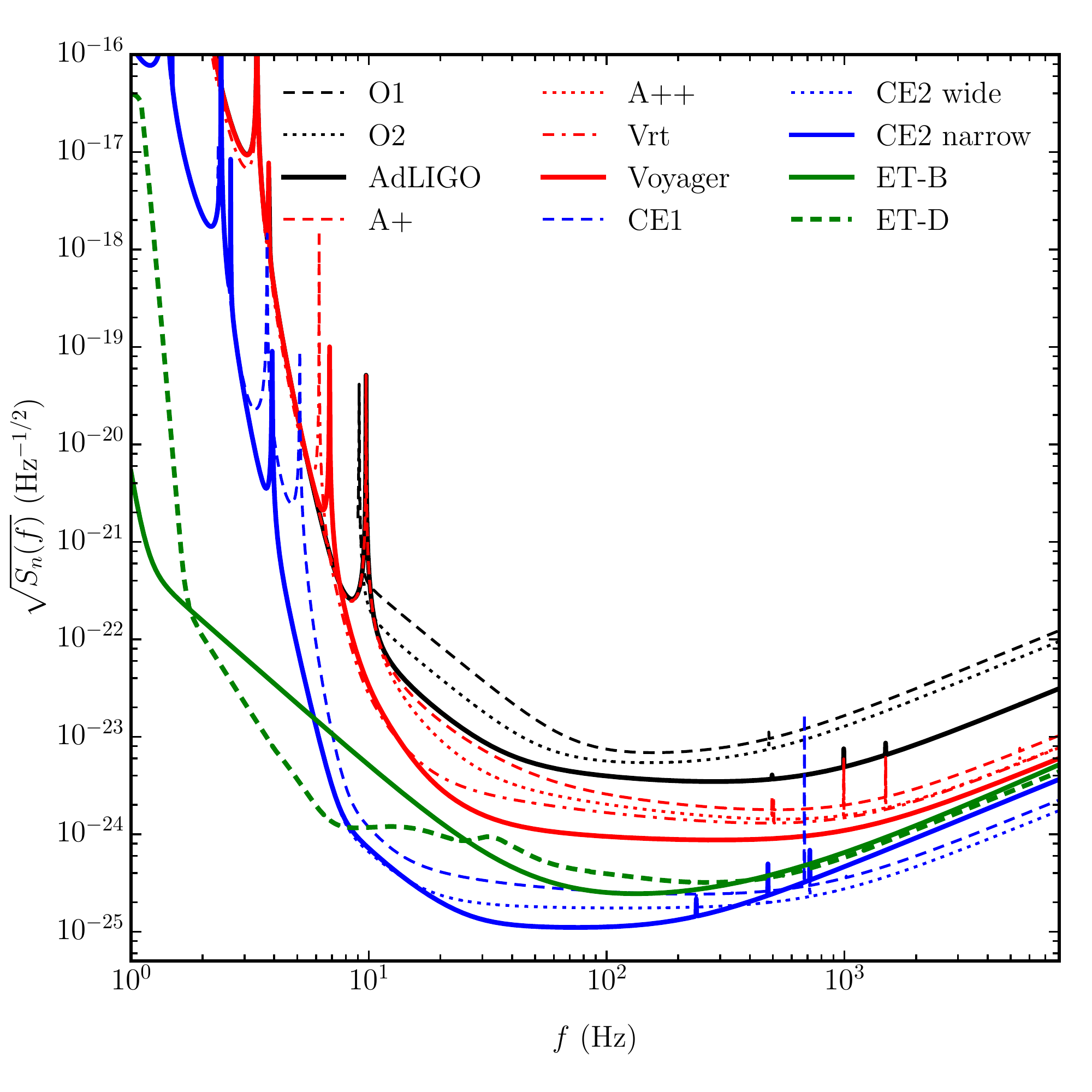}
	\caption{Noise power spectral densities (PSDs) for present and
          future Earth-based detector designs (see Sec.~\ref{sec.rate} for 
          details).}
	\label{fig.noise}
\end{figure}
 
\begin{figure}
  \begin{center}
	\includegraphics[width=\columnwidth]{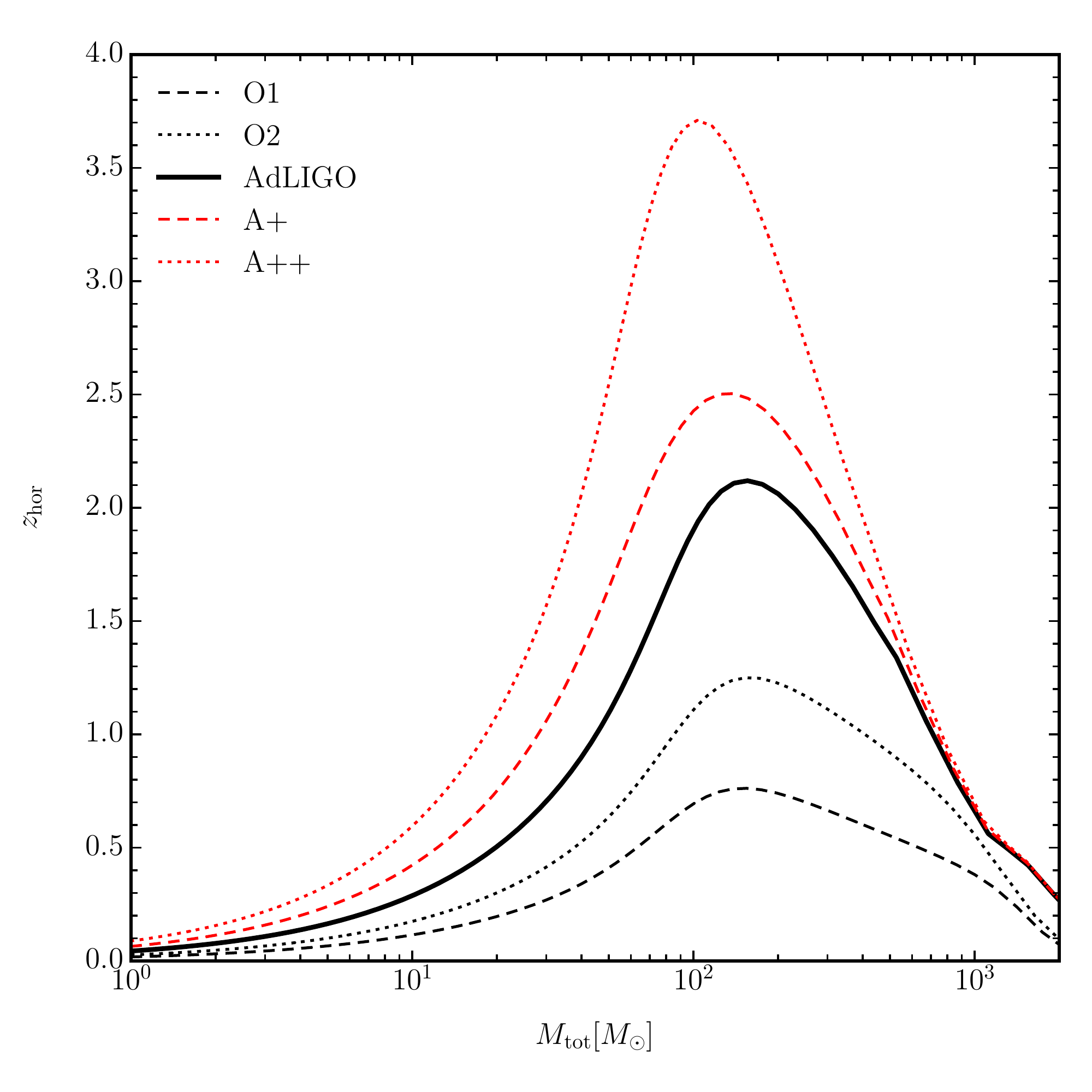}
	\includegraphics[width=\columnwidth]{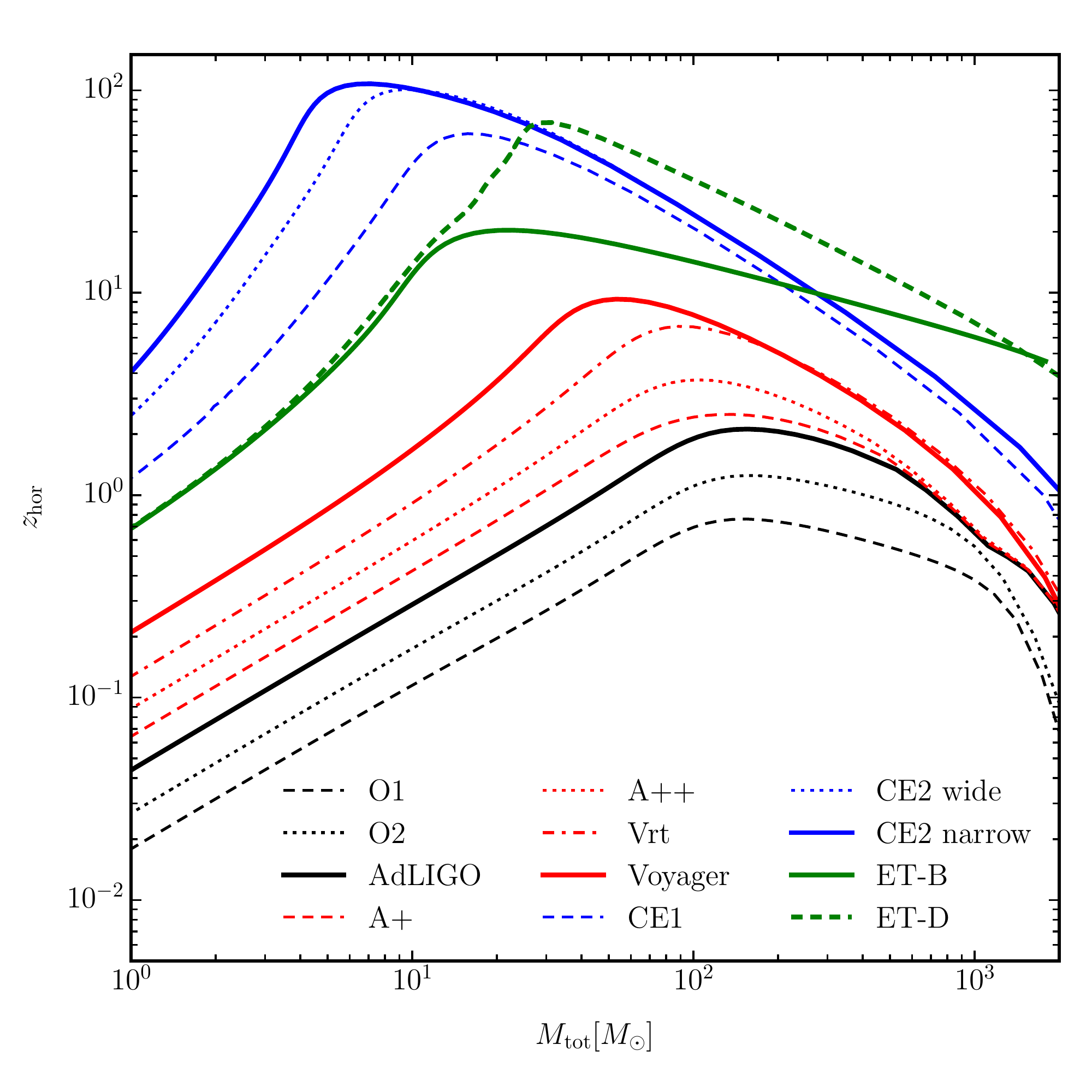}
  \end{center}
	\caption{
		Horizon redshift for various detectors. 
	}
	\label{fig.horizon}
\end{figure}

\subsection{BH-BH delay time}

In Figure~\ref{fig.del} we show the delay time distributions for our models of 
Pop~III stars (FS1 and FS2). The delay time includes evolutionary
time from star formation, to the formation of BH-BH binary, and the coalescence
time to the final merger. 
The delay time distribution for model FS1 is a steep power-law: for very
short delay times ($t_{\rm del}<0.5$ Gyr) the number of BH-BH mergers falls 
off as $\propto t^{-3}$, and for longer delay times it falls of a bit faster 
than $\propto t^{-1}$. 

The delay time is the result of a convolution of the initial separation
distribution that gets modified by stable mass transfer and CE during evolution 
(see Tab.~\ref{tab.evol}) and the timescale for orbital decay due to emission of
gravitational waves once the BH-BH binary is formed. Stable mass transfer does not 
modify the  orbital size significantly, while CE leads to significant contraction. 
For massive BH progenitors, the CE contraction reduces the orbital size
typically by a factor of about $100$, and does not significantly change
the shape of the distribution of orbital separations. 

For initial separations that are flat or approximately flat in
logarithm ($\propto a^{-1}$) in the range of BH-BH formation
($a\gtrsim1000\rsun$), the convolution with gravitational radiation
emission orbital decay timescale ($\propto a^4$; \citealt{Peters1964})
results in power-law delay time distribution $\propto t^{-1}$, because
the delay time scales like
$a^{-1} (da/dt)_{\rm GR} \propto t^{-1/4} d(t^{1/4})/dt \propto
t^{-1}$).  In fact, for Pop~I/II binaries BH-BH delay times follow
such a power-law $\propto t^{-1}$ \citep[e.g][]{Belczynski2016b}.
However, since Pop~III BH-BH mergers host BHs with larger masses (see
Fig.~\ref{fig.totmass}) than Pop~I/II mergers, the delay time
becomes steeper for Pop~III mergers. The coalescence time is
proportional to $[M_1M_2 (M_1+M_2)]^{-1}$ \citep{Peters1964}.

\subsection{BH-BH Merger Rate Density}
\label{sec.rate}

In Figure~\ref{fig.rates} we show the rate density of BH-BH mergers from
our Pop~III models, FS1 and FS2. For comparison we also show 
the results for our population synthesis model for Pop~I/II stars, M10. 

The merger rate density is a combination of star formation rate density (see
Fig.~\ref{fig.sfr}), BH-BH merger formation efficiency (see Tab.~\ref{tab.eff}) 
and time delay from formation of stars to BH-BH merger (Fig.~\ref{fig.del}). 

Since the delay times are in a form of steep (negative) power-laws for all
our models (most systems have relatively short delay times) the evolution 
of the BH-BH merger rate density with redshift has a shape similar to the star 
formation rate density appropriate for each stellar population. For Pop~III model 
FS1, the rate density starts at high redshift ($z\approx45$), and then
increases to its peak at $z\approx10$--$13$, and then decreases for smaller
redshifts. In the redshift range of interest for advanced LIGO ($z<2$), the rate
density is at the level of $R_{\rm BHBH}\approx 0.1 \gpy$.  
 
In Figure~\ref{fig.noise} we show the noise power spectral densities
(PSDs) for present and future Earth-based detector designs. From top
to bottom, the figure shows the sensitivity of the first Advanced LIGO
observing run (O1); the expected sensitivity for the second observing
run (O2); the Advanced LIGO design sensitivity (AdLIGO)~\citep{LVC2013}; 
pessimistic and optimistic ranges of Advanced LIGO designs with squeezing 
(A+, A++)~\citep{Miller2015}; the most sensitive interferometers that can be 
built in current facilities, Vrt and Voyager~\citep{Adhikari2014,LIGONextGeneration2016}; 
Cosmic Explorer (CE1), basically A+ in a 40-km facility~\citep{Dwyer2015}; CE2 wide
and CE2 narrow, i.e. 40-km detectors with Voyager-type technology but
different signal extraction tuning~\citep{LIGONextGeneration2016}; and
two possible Einstein Telescope designs, namely ET-B\footnote{See
  \url{http://www.et-gw.eu/etdsdocument}.}
and ET-D in the ``xylophone'' configuration~\citep{Hild2010}.

In order to compare the detectability of Pop~I/II and Pop~III sources,
a particularly interesting quantity is the horizon redshift
$z_{\rm hor}$. We define $z_{\rm hor}$ as the redshift out to which
any given detector can observe the full inspiral-merger-ringdown
signal from an optimally oriented, nonspinning BH binary. 
\cite{Dominik2015} showed that spin effects may increase BH-BH
detection rates by as much as a factor of 3, but for simplicity we
will ignore spins in our model comparisons. The maximum horizon
redshift is achieved for equal-mass binaries, and it is shown in
Figure~\ref{fig.horizon} as a function of the total intrinsic mass $M_{\rm tot}$ 
of a nonspinning BH binary (as measured in the source frame). For this
calculation we estimate the gravitational wave signal using the phenomenological
PhenomC waveforms of \cite{PhenomC}.  The sensitivity of a gravitational wave detector
network depends on the details of the search pipeline and the detector
data quality, but following \cite{LIGORates} we set a single-detector
SNR threshold $\rho\geq 8$ as a proxy for detectability by the
network. The left panel shows that binaries of source mass $\approx
10^2 M_\odot$ can be detected out to $z_{\rm hor}\approx 1$ in
the second observing run O2, $z_{\rm hor}\approx 2$ by AdLIGO at
design sensitivity, and $z_{\rm hor}\approx 4$ by an Advanced LIGO
detector with squeezing in the ``optimistic'' (A++) configuration. The
right panel shows that detectors such as Voyager or ET could reach
redshifts of order 10, and that detectors of the ``Cosmic Explorer''
class would see gravitational waves from binaries throughout the whole Universe.

From Figure~\ref{fig.rates} we see that BH-BH mergers first occur
around $z\approx15$ for the Pop~I/II model M10; the merger rate
density peaks at $z\approx2$, and decreases for smaller redshifts. In
the LIGO range ($z<2$) the rate density is at the level of
$R_{\rm BHBH}\approx 1000 \gpy$. Note that the Pop~I/II merger rate
density is about four orders of magnitude larger than for
Pop~III. This is a direct result of the fact that (i) Pop~I/II star
formation produces many more stars ($99.7\%$) than Pop~III ($0.3\%$),
(ii) Pop~I/II star formation peaks within the Advanced LIGO horizon
($z\approx2$), while Pop~III stars form mostly well outside the
Advanced LIGO horizon ($z\approx10 - 13$), (iii) the delay times from
star formation to BH-BH mergers are short for both Pop~I/II and
Pop~III mergers ($\propto t^{-1}$), and (iv) the formation efficiency
of BH-BH mergers per unit of star forming mass is similar
($10^{-5} \msun^{-1}$ and $10^{-4} \msun^{-1}$ for Population I/II and
Pop~III, respectively).

Even before computing rates, we can already attempt a rough comparison
with existing LIGO data. In the local Universe ($z\lesssim0.1 - 0.2$)
LIGO observations have delivered the following constraint on the BH-BH
merger rate density: $R_{\rm BHBH}=9 - 240\gpy$~\citep{LigoO1b}.  The
Pop~I/II model M10 of classical isolated binary evolution is close to
the upper end of this range, with $R_{\rm BHBH}=220\gpy$ (see
Table~\ref{tab.det}). By contrast, our Pop~III model FS1 yields
$R_{\rm BHBH}=0.02\gpy$ (see Tab.~\ref{tab.det}), $\sim 2.5$ orders of
magnitude below the lower limit of the current LIGO estimate, and
$4$ orders of magnitude below the Pop~I/II estimate. These
expectations are confirmed by a detailed calculation of detection
rates for current and future detectors, as shown below.

\begin{table}
	\caption{Merger rate densities and detection rates for AdLIGO}             
	\label{tab.det}      
	\centering
	\begin{threeparttable}
	\begin{tabular}{c| c c c}     

\hline\hline       
model & rate density\tnote{a} & det. rate\tnote{b} & det. number\tnote{c}\\ 
merger type   & [Gpc$^{-3}$ yr$^{-1}$] & [yr$^{-1}$] & [131 days] \\ 
\hline          
   FS1 &&&\\         
   NS-NS &          0$\to$0 &     0  &     0  \\  
   BH-NS &      0.002$\to$0 & 0.012  & 0.004  \\
   BH-BH &  0.022$\to$0.234 & 5.974  & 2.151  \\
\hline
   FS2 &&&\\
   NS-NS &   0$\to$0  & 0   & 0  \\  
   BH-NS &   0$\to$0  & 0   & 0  \\
   BH-BH &   0$\to$0  & 0   & 0  \\
\hline                  
   M10 &&&\\
   NS-NS &  74.0$\to$121 & 1.418 & 0.510 \\  
   BH-NS & 27.2$\to$84.2 & 6.951 & 2.502 \\
   BH-BH &  220$\to$1802 &  2080 & 748.8 \\
\hline
   KK1 &&&\\
   NS-NS &         0$\to$0 &     0 &     0 \\  
   BH-NS &     0.004$\to$0 & 0.020 & 0.007 \\
   BH-BH & 0.024$\to$0.153 & 4.067 & 1.464 \\
\hline
   KK2 &&&\\
   NS-NS &        0$\to$0 &     0 &     0 \\  
   BH-NS &        0$\to$0 &     0 &     0 \\
   BH-BH & 1.159$\to$12.0 &   507 &   182 \\

\hline \hline                  
		
	\end{tabular}
	\begin{tablenotes}
		\item[a] 
Typically, merger rate density increases from low redshift to high redshift 
within advanced LIGO horizon. We list rate density at $z=0$ (local; before 
arrow) and at $z=2$ (AdLIGO horizon; after arrow). 
	    \item[b] Detection rate for full advanced LIGO sensitivity.
	    \item[c] Number of LIGO detections per 1 years of advanced
              LIGO, while assuming $p=0.36$ duty cycle: $131$
              effective observation days per year. 
		\end{tablenotes}
		\end{threeparttable}
\end{table}

\subsection{Advanced LIGO Detection Rates}
\label{sec.detection}

Advanced LIGO is expected to achieve design sensitivity around
2020. Following \cite{Dominik2015}, we use the ``AdLIGO'' design
sensitivity in Figure~\ref{fig.noise} and the PhenomC model for
nonspinning BH binary merger waveforms to estimate the number of
detected events for both Pop~III and Pop~I/II evolutionary models. 

The results of this calculation are presented in Table~\ref{tab.det}.
According to model M10, Pop~I/II BH-BH mergers are expected to be
detected at a rate of $\sim 2000$ yr$^{-1}$. However we should take
into account the fact that the LIGO interferometers will not work
simultaneously all the time. The typical duty cycle $p$ in the first
observing run O1 (i.e., the fraction of time during the run when both
interferometers were operating simultaneously) was about $p=0.36$.
Taking into account the duty cycle, the predicted rate yields about
$\sim 700$ detections in one year of actual observations.  These
detection rate estimates are on the high side for Pop~I/II stars;
other models compatible with current LIGO constraints predict smaller
rates. Rates that are smaller by about one order of magnitude cannot 
be excluded \citep{Belczynski2016c}. Such reduced rates of BH-BH 
mergers are obtained for example with increased BH natal kicks. In
particular, even very high BH natal kicks cannot yet be excluded based 
on electromagnetic observations~\citep{Repetto2015,Belczynski2016a,Mandel2016b}. 

By comparison, the Pop~III BH-BH AdLIGO detection rate is much lower: for 
model FS1 we find a rate of $\sim 6$ yr$^{-1}$, or $\sim 2$ yr$^{-1}$ if we
take into account the duty cycle of the detectors. These numbers are
{\it only} $2.5$ orders-of-magnitude lower than the corresponding
numbers for model M10. The BH-BH merger rate density difference
between the two models is much larger: $4$ orders-of-magnitude (see
Sec.~\ref{sec.rate}). This is explained by the higher mass of
merging BH-BH binaries formed from Pop~III stars
($M_{\rm tot}=63.4\msun$; see Fig.~\ref{fig.totmass}) as compared
with mergers from Pop~I/II stars ($M_{\rm tot}=29.7\msun$). More massive 
binaries emit stronger gravitational waves, and therefore they have a larger
detector horizon (see Fig.~\ref{fig.horizon}). Therefore, Pop~III
mergers can be seen at distances about twice as large (corresponding
to a volume about $10$ times as large). This is why the ratio of
detection rates between Pop~I/II to Pop~III mergers is lower than
the ratio of their intrinsic merger rate densities.

Detection rates for Pop~III BH-BH mergers are small but non-zero. Our
results suggest that when Advanced LIGO attains design sensitivity, a
handful of Pop~III mergers may be expected in a population of tens or
hundreds of BH-BH mergers originating from Pop~I/II stars. This small
contribution ($\lesssim 1\%$) may go unnoticed and remain hidden in
the population of other BH-BH mergers. Note that Pop~III BH-BH mergers 
reach the same maximum intrinsic mass as Pop~I/II BH-BH mergers ($M_{\rm
tot,max} \approx 80 - 90 \msun$; see Fig.~\ref{fig.totmass}). The maximum
total mass is set, for both Pop~III and Pop~I/II, by pair instability
pulsation supernovae mass loss and by binary evolution that removes H-rich
envelopes from stars that form BH-BH mergers (see Sec.~\ref{sec.bhmass} and 
~\ref{sec.evolution} for details). Although the exact value of the total 
maximum mass may be uncertain, the same cutoff is expected for both Pop~III
and Pop~I/II BH-BH mergers. Therefore, Pop~III BH-BH mergers are not 
likely to be distinguished through their heavier mass by AdLIGO. 
Although the distribution of total mass for Pop~III BH-BH mergers is 
different from that of Pop~I/II BH-BH mergers (see Fig.~\ref{fig.totmass}), 
the very small number of predicted AdLIGO detections for Pop~III BH-BH 
mergers will not allow accurate (if any) measurement of their mass 
distribution. These Pop~III BH-BH mergers will most likely blend into 
larger populations of Pop~I/II BH-BH mergers. 

This is in tension with the conclusion reached by \citet{Hartwig2016} that
gravitational waves have the potential to directly detect the remnants of 
the first stars, and possibly even to constrain the Pop~III IMF by observing 
BH–BH mergers with a total mass around $M_{\rm tot}=300\msun$.
This different conclusion originates from the fact that \citet{Hartwig2016} 
considered a Pop~III IMF that may potentially extend to 
$M_{\rm zams}=300\msun$ or even to $M_{\rm zams}=1000\msun$. This produces 
a small but distinctive population of Pop~III BH-BH mergers with 
total mass of about $M_{\rm tot}=300\msun$, which would stand out from 
Pop~I/II BH-BH mergers simply by the virtue of its high total mass. This 
is only true if Pop~I/II stars have IMF that do not extend above a mass of 
$M_{\rm zams}=300\msun$. However, there is observational evidence that 
Pop~I/II stars may reach such high mass: several stars in the R136 region 
of the Large Magellanic Cloud may have had an initial mass up to 
$300\msun$~\citep{Crowther2010}. Also, recent detailed evolutionary models 
have shown that stars as massive as $500\msun$ can exist and evolve at 
low- to moderate-metallicity, forming massive BHs \citep{Yusof2013}. 
In fact, for $Z=0.002$ (about $10\%$ solar metallicity) very massive stars 
can lead to formation and AdLIGO detection of several to hundreds of BH-BH 
mergers within a total intrinsic mass range $M_{\rm tot}=50 - 300\msun$ (see 
Fig.3 of~\citealt{Belczynski2014}). 
If the metallicity were decreased to say $Z=0.0002$ (about $1\%$ solar 
metallicity), the total intrinsic mass for these BH-BH mergers would increase.
This would cause the overlap of BH-BH massive mergers from Pop~I/II very
massive stars with BH-BH massive mergers from Pop~III stars. Whether such
very massive stars exist ($M_{\rm zams} \gtrsim 300\msun$; as to avoid pair
instability supernova disruption and allow for massive BH formation) in 
Pop~III and in Pop~I/II stars still remains an open question. However, if 
such stars are considered in one population, they should also be considered 
in the other population. Therefore, the idea that Pop~III stars can be
investigated by means of massive BH-BH mergers needs to allow for this caveat. 

We note that, while our predicted rates are considerably lower than those 
found by other investigations (see Sec.~\ref{sec:KinugawaComparison} for a 
detailed model comparison), they are consistent with those estimated by 
\citet{Hartwig2016}. These authors used a combination of dark matter halo 
merger trees (GALFORM; \citealt{Parkinson2008}) specific for Pop~III star
formation with some (e.g., initial binary separations) properties specific 
for massive Pop~I stars~\citep{Sana2012} to determine the initial conditions 
for estimates of BH masses and BH-BH merger rates. These estimates were based 
on population synthesis study for Pop~I stars with $Z=0.002$: with non-zero 
wind mass loss and significant radial expansion of stars and without effects 
of pair-instability pulsation supernovae~\citep{deMink2015}. 
\citet{Hartwig2016} Pop~III SFR is significantly lower than the rate assumed 
by \citet{Kinugawa2014}; the latter rates come from the model of 
\citet{deSouza+11}, which we have also used in this work (see Fig.~\ref{fig.sfr}).  

The model of \citet{Hartwig2016} predicts up to $5.3$ detections per year of 
Pop~III BH-BH mergers by AdLIGO. This is remarkably similar to our result: 
$6.0$ detections per year (see model FS1 in Tab.~\ref{tab.det}). 
Modifications of Pop~III IMF lead \citet{Hartwig2016} to decreased rates: 
down to $0.12$ BH-BH detections per year. The similarity of results is the 
effect of the fact that the low Pop~III SFR was compensated by the effective 
production of BH-BH mergers in the population synthesis calculations: high 
binary interaction rates resulting from relatively short initial binary 
separations and significant stellar radial expansion employed in 
\citet{Hartwig2016}.

\begin{figure}
  \begin{center}
	\includegraphics[width=8cm]{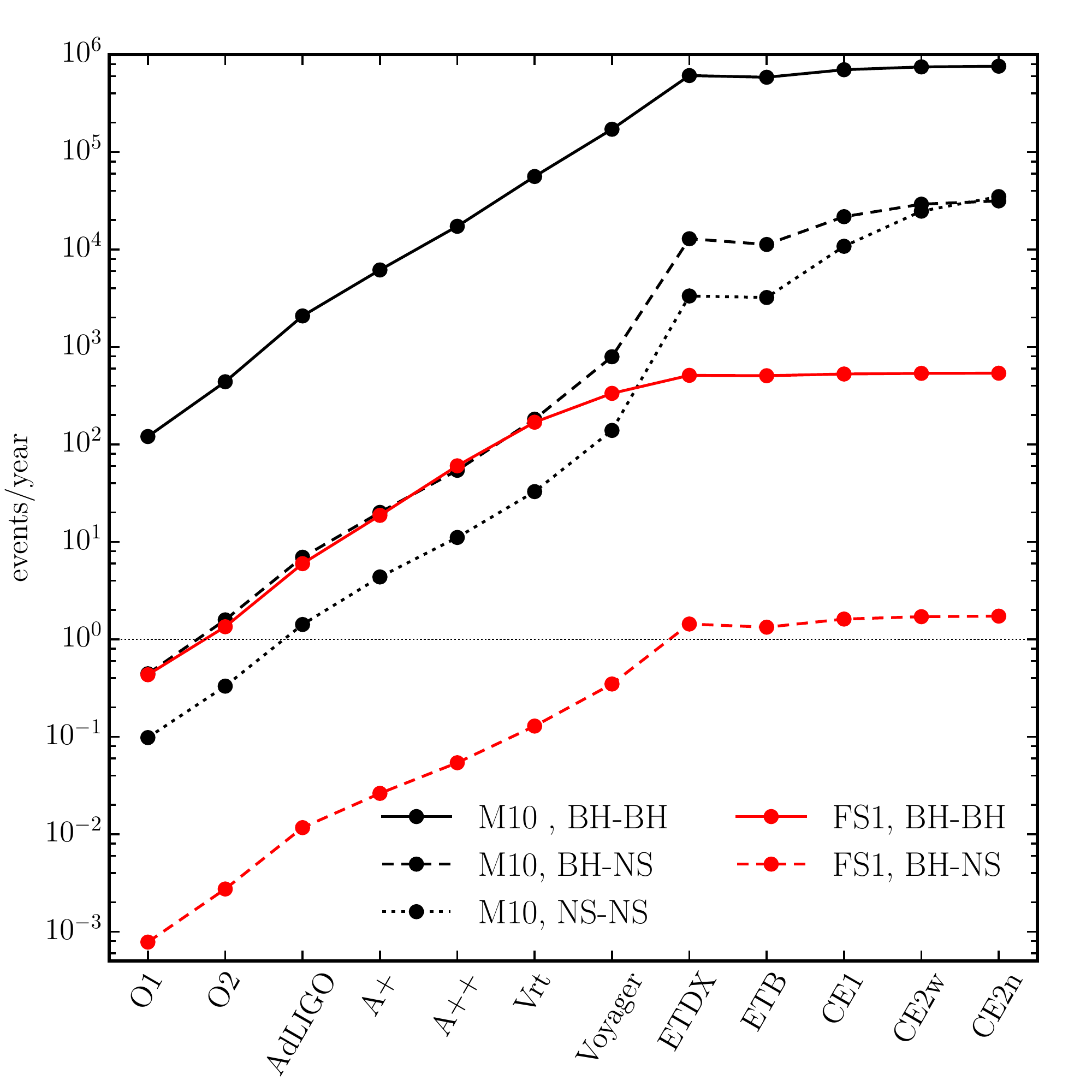}
  \end{center}
  \caption{ Rates for various detectors. The horizontal bar marks the
    threshold of one detection per year, above which observations
    become likely.}
	\label{fig.detrates}
\end{figure}

In Figure~\ref{fig.detrates} we show a more comprehensive calculation
of detection rates for all three classes of compact binaries according
to models M10 and FS1, including all types of double compact objects: 
BH-BH, BH-NS and NS-NS mergers. The rate calculation follows closely the 
method described in \cite{Dominik2015}. Unlike Table~\ref{tab.det}, where 
we listed results only for Advanced LIGO at design sensitivity, here we 
show event rates for all of the noise PSDs shown in Figure~\ref{fig.noise}. 

In the Pop~I/II model M10, BH-NS and NS-NS detection rates are comparable 
in order of magnitude (BH-NS rates being slightly larger for all detectors); 
BH-BH detection rates are about two orders of magnitude higher. While a BH-NS 
observation may occur already during the O2 run, NS-NS detections seem to be 
likely only at the Advanced LIGO design sensitivity and above. 

Our most optimistic Pop~III model (FS1) predicts BH-BH rates comparable to the 
BH-NS rates in model M10, because the high mass of Pop~III BHs compensates for 
their lower merger rate density. The number of detections for Pop~III BH-BH
mergers plateaus for third generation detectors such as Voyager, ET and Cosmic
Explorer, essentially because all detectable binaries in the Universe
have already been seen, and better noise PSDs only increase the SNR of
the observed events; by contrast, the number of BH-NS and NS-NS detections in 
the M10 model keeps increasing as detectors are improved and more of the 
Universe becomes visible.

\begin{figure}
  \begin{center}
	\includegraphics[width=8cm]{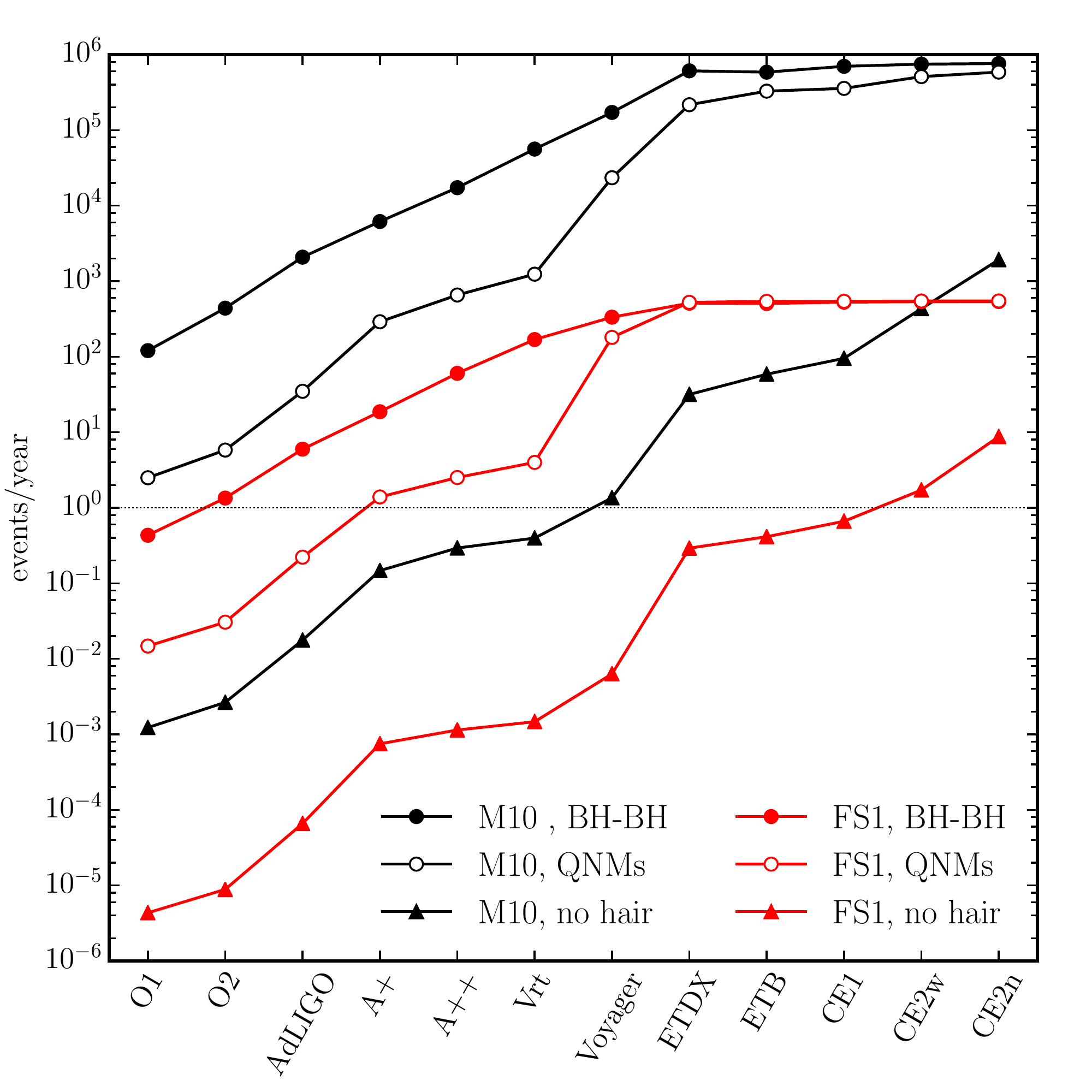}
  \end{center}
  \caption{Full inspiral-merger-ringdown rates (solid circles),
    ringdown detection rates (empty circles) and number of detections
    allowing us to do BH spectroscopy (triangles) for models M10
    (black) and FS1 (red), and for various detectors.}
	\label{fig.spectroscopy}
\end{figure}

\subsection{Binary black hole spectroscopy}

After merger, two BHs settle down to a stationary (Kerr) solution of
Einstein's equations by emitting gravitational waves at characteristic complex
frequencies, known at the ``quasinormal mode frequencies''
(``quasinormal'' because the system is dissipative and the frequencies
are complex: the inverse of the imaginary part corresponds to the
damping time due to gravitational wave emission). In GW150914, this ``ringdown''
signal was observed with a relatively low SNR $\rho\sim 7$. The signal
is consistent with the predictions of general relativity for a Kerr
BH, but it does not allow us to conclusively confirm the Kerr nature
of the remnant. As first pointed out by \cite{Detweiler1980}, with
higher-SNR observations we may be able to measure more than one
quasinormal mode frequency. Since all quasinormal mode frequencies in
general relativity depend only on the BH mass and spin, a multi-mode
gravitational wave detection would allow us to do ``BH spectroscopy'' -- i.e., to
identify these objects as Kerr BHs with the same certainty with which
we identify atoms from their spectra \citep{QNMReview}. To quote
Detweiler: ``after the advent of gravitational wave astronomy, the
observation of these resonant frequencies might finally provide direct
evidence of BHs with the same certainty as, say, the 21 cm line
identifies interstellar hydrogen.'' As discussed by \cite{BCW} and
\cite{BCCC}, the detection of a second mode requires SNRs ($\gtrsim 80$) 
about one order of magnitude larger than the SNRs ($\gtrsim 8$) necessary 
for ringdown detection.

In Figure 14 we consider, for models M10 and FS1, the rates per year
of BH-BH mergers for which we can detect the full
inspiral-merger-ringdown signal (solid circles) and the ringdown
signal (empty circles). We also show the (much smaller) number of
detections that would allow us to do BH spectroscopy (triangles).
Detectable BH ringdown signals and rates for BH spectroscopy are
estimated as described in \cite{Berti2016}. Because BH-BH merger event
rates in model FS1 are two orders of magnitude lower than in model
M10, ringdown detections and spectroscopic tests are also much more
unlikely. According to our model, the detection of ringdown signals
from Pop~III binaries will require at least the implementation of
squeezing.  Notice that while BH merger (and ringdown) detections
plateau for detectors such as ET or Cosmic Explorer and beyond, the
fraction of events allowing for BH spectroscopy increases: this is
because the SNR of the observed events is increasing, and no-hair
tests ultimately become possible with 40-km facilities such as CE2w.

\section{The origin of the unphysically high Pop~III BH-BH merger rates}
\label{sec:KinugawaComparison}

Since the Pop~III rates that we find here are considerably lower than those 
found by other studies (i.e. \citealt{Bond1984b,Belczynski2004,Kinugawa2014}),
we have devoted a considerable effort to understand the reasons for these 
differences, and performed a detailed model comparison, as described below.
We however note that our detection rates of Pop~III BH-BH mergers are in 
agreement with those recently found by \citet{Hartwig2016} 
(see Sec.~\ref{sec.detection}).

\subsection{Description of the models used for a comparative study}

We use one of the recent studies that argues for high Pop~III BH-BH 
merger rates and significant contribution of these mergers to advanced LIGO 
signal to show that such findings originate from {\textit (i)} initial conditions
that may not be appropriate for Pop~III stars,
 and {\textit (ii)} outdated evolutionary calculations for Pop~III
binaries.  In particular, \cite{Kinugawa2014} finds a merger rate density
of $12$--$25\gpy$ at small redshifts (see their Fig.9) and an advanced LIGO detection 
rate at the level of $68$--$140$ yr$^{-1}$ for Pop~III BH-BH mergers. 
The range of the reported rates corresponds to different assumption for the Pop~III IMF. 
The low rates correspond to Salpeter-like IMF (power-law with index
of $-2.3$), while the high rates correspond to flat IMF. We will use the
high rate model of \cite{Kinugawa2014} for our comparison, and hence we adopt
the flat IMF. 

We constructed two models in order to reproduce and explain the results of 
\cite{Kinugawa2014}. 
In model KK1 we employ initial conditions for evolution of Pop~III
binaries as in \cite{Kinugawa2014}, \textit{but} we keep our evolutionary scheme. 
In model KK2 we employ initial conditions for evolution of Pop~III   
binaries as in \cite{Kinugawa2014}, \textit{and} we match (as closely as we can) 
the evolutionary scheme of \cite{Kinugawa2014}. For both models we use the same
SFR used by \cite{Kinugawa2014}, which we have also adopted for our 
models (FS1 and FS2; see Fig.~\ref{fig.sfr}). Also the radius evolution of the 
Pop~III stars in our models matches closely that of \cite{Kinugawa2014} (see 
Fig.~\ref{fig.radius}). 

Model KK1 initial distributions are as follows. The primary mass ($M_1$) is drawn 
from a flat IMF within the range $10$--$100\msun$. The mass ratio (secondary-to-primary) 
is taken from a flat distribution in the range $q_{\rm min} - 1$ with 
$q_{\rm min}=10/M_1$. For the eccentricity distribution we adopt $f(e)=2e$ in 
the range $0 - 1$, while  for the orbital separation distribution we adopt $f(a)=1/a$ in
range $a_{\rm min} - 10^6\rsun$, with $a_{\rm min}$ such that stars do not
overfill their Roche lobes at periastron on Zero Age Main Sequence. The binary
fraction is taken to be $f_{\rm bi}=1/3$. We evolve the stars as described in
Sec.~\ref{model}.

Model KK2 uses the same initial distributions as in model KK1. However, the evolution
is modified in the following way. 

\begin{itemize}  
\item[(i)] We adopt a constant $\lambda=1.0$: this is a parameter that describes CE 
binding energy. Note that more realistic values used in our current simulations 
(models FS1, FS2, M10) are not constant and are much smaller ($\lambda \sim 0.1$; see 
Sec.~\ref{sec.evolution}) so the binding energy in our models is much higher 
($E_{\rm bind} \propto \lambda^{-1}$).

\item[(ii)] We use our old computation scheme of BH and NS mass
\citep{Belczynski2002} that was obtained for outdated wind-mass loss
prescriptions for massive stars. The new (weaker) mass loss
\citep{Vink2001,Belczynski2010b,Vink2011} and new supernova models 
(\citet{Fryer2012}; rapid explosions) were applied in our current 
simulations (models FS1, FS2, M10).

\item[(iii)] We do not apply corrections that take into account pair-instability
supernovae and pair-instability pulsation supernovae. Note that these supernovae 
disrupt the most massive stars and reduce the BH mass for high mass stars (see 
Fig.~\ref{fig.bhmass}). 

\item[(iv)] We do not apply natal kicks, neither to NSs nor to BHs. In all
our other models considered in this study, compact objects receive natal 
kicks based on the formation mode. 
Light NSs receive average 3D kicks of about $400$ km s$^{-1}$ \citep{Hobbs2005}, 
the most massive BHs that form through direct collapse get no natal kicks, while 
the heavy NSs and light BHs receive natal kicks decreasing with increasing mass of 
fall back material \citep{Fryer2012}. 

\item[(v)] We allow stars on HG (radiative envelopes) and beyond to enter and 
survive CE. This allows for enhanced formation of merging double compact objects.
This is in contrast with what we apply in other models; only stars beyond HG 
during later stages of evolution (convective envelopes) are allowed to survive CE 
(see Fig.~\ref{fig.radius}). 

\item[(vi)] Additionally, we turn off magnetic braking and set the maximum NS mass
at $3\msun$ (instead of $2.5\msun$ used in our other models).

\end{itemize}

\subsection{Comparison with Kinugawa et al. 2014} 

The results for models KK1 and KK2 are given in Table~\ref{tab.det} (merger
rate densities and detection rates), Figure~\ref{fig.ratesK} (BH-BH merger 
rate density change with redshift), Figure~\ref{fig.totmassK} (BH-BH merger 
total mass) and in Figure~\ref{fig.delK} (BH-BH merger delay time). 

We note that our model KK2 resembles rather closely that of \cite{Kinugawa2014}
(one with flat IMF; noted in their work as model III.f). The KK2 BH-BH merger 
rate density increases from $1.2\gpy$ ($z=0$) to $12\gpy$ ($z=2$) within reach of
advanced LIGO. For comparison, the  reported BH-BH merger rate density in model 
III.f is $25\gpy$ (a factor of $\sim 2$ larger than our maximum rate density 
within the LIGO horizon).  

The KK2 BH-BH detection rate for advanced LIGO is $507$ yr$^{-1}$,
while the detection rate in model III.f by \cite{Kinugawa2014} is only
$182$ yr$^{-1}$.  This is surprising, since the average intrinsic
total mass of BH-BH mergers in model KK2 ($M_{\rm tot}=62.8\msun$; see
Fig.~\ref{fig.totmassK}) is rather similar to the one in model III.f
($M_{\rm tot} \approx 55\msun$; read off Fig.6 of
\citealt{Kinugawa2014}).  This may result from an improper calculation of
the advanced LIGO rate.  The estimates by \citet{Kinugawa2014} account
only for sources up to the redshift of $z=0.28$, while the advanced
LIGO design sensitivity will allow to detect massive BH-BH mergers to
$z\approx2$ (see Fig.~\ref{fig.horizon}). Additionally, no signal waveforms 
nor the projected advanced LIGO response function are used in their 
calculations of the detection rates; they simply count merging sources within 
the redshift of $z=0.28$ (see their eq.~95). Hence in the following we will 
only make comparisons with their merger rate density, and ignore
\cite{Kinugawa2014} estimate of the detection rate.

The factor of $\sim 2$ difference in the BH-BH merger rate density
noted between models KK2 and III.f originates from the fact that
\cite{Kinugawa2014} includes a BH-BH formation channel that is in
tension with evolutionary studies done so far for binary stars in
isolation and without rapid rotation \citep{Tutukov1993,Lipunov1997,
Belczynski2002,Voss2003,Postnov2006,Belczynski2010a,Dominik2012,
Mennekens2014,Belczynski2016b,Eldridge2016}. 
Note that \cite{Kinugawa2014} does not deal
with dynamical BH-BH formation, and that their code is based on
updated \cite{Hurley2000} formulas (same as our model KK2) so they
cannot model rapidly rotating stars and homogeneous evolution.
Therefore, their evolutionary scenario would need to have a common
envelope stage in order to bring two massive BHs within a distance
small enough so that they can merge within a Hubble time. For two
$30\msun$ BHs (their typical mass of Pop~III BH-BH mergers) on a
circular orbit one requires the orbital separation to be below
$50\rsun$ for the system to merge within Hubble time. Two massive
stars with $M_{\rm zams}=40-100\msun$ \citep{Belczynski2016b}, progenitors
of such massive BHs, cannot fit on an orbit with a size below
$50\rsun$. So typically, they start on wide orbits ($1000-4000\rsun$;
\citealt{deMink2015}) and then CE brings the compact remnants
together. There is no published alternative for such a formation
scenario for isolated and slow-to moderately-spinning stars. And yet,
their major channel ($36.9\%$ of BH-BH mergers; the second most
populated channel is only $16.3\%$: see their Table 5) does not
require common envelope. The details of this evolutionary path hence 
remain unclear. If we
look at the results of their simulations without this channel, then our
results in model KK2 are very close to their model III.f.  In
Table~\ref{tab.eff} we show that each channel of our evolution (and in
particular model KK2) requires a common envelope phase.

In addition to these differences in the evolutionary scenarios, the
main reason for the discrepancies in the predicted merger rates lies
in the adopted initial conditions. \citet{Kinugawa2014}, and likewise our
models KK1 and KK2, use thermal distribution of eccentricities and
flat in log orbital separation distribution for their initial
conditions. These are outdated conditions that were used in the early
stages of population synthesis codes for Pop~I/II stars
\citep{Belczynski2002}. Additionally, it has now been demonstrated that
these particular distributions do not apply even to massive Pop~I stars 
\citep{Kobulnicky2007,Sana2012,Chini2012,Kobulnicky2014,Moe2016}. 
However, they are crucial for the production
rates of BH-BH mergers \citep{deMink2015}. As we have shown here
(models FS1 and FS2), these initial distributions are expected to be 
very different for Pop~III stars.
Additionally, \citet{Kinugawa2014} assume unrealistic conditions for
CE, which is again crucial for BH-BH merger production. More specifically,
they assume a very low binding energy for the envelope of massive
stars ($\lambda=1.0$), which is in tension with all the recent work on
the subject \citep{Xu2010,Loveridge2011,Kruckow2016}.

Other recent estimates of the stochastic gravitational wave background
indicate a relatively high (even dominating) contribution of Pop~III
BH-BH mergers \citep{Kowalska2012,Inayoshi2016,Dvorkin2016}. These
results stem, like in \citet{Kinugawa2014}, from the use of Pop~I
specific initial conditions for Pop~III binaries and/or an outdated
evolution framework for estimates of the BH-BH merger formation
efficiency in Pop~III binaries (see
Sec.~\ref{sec:KinugawaComparison}).  However, if initial conditions
more specific to Pop~III binary formation and evolution were rather
used, it is likely that the stochastic gravitational wave background
from Pop~III BH-BH mergers may become insignificant. We provide data
on our Pop~III BH-BH mergers (model FS1) at
\url{www.syntheticuniverse.org}.  These data can be used for what we
believe are more realistic estimates of the background from Pop~III
BH-BH and BH-NS mergers.

To summarize, if more appropriate Pop~III initial conditions are
adopted, and an updated evolutionary scheme is used, the BH-BH merger
rates fall below $0.1\gpy$ at low redshifts (see Fig.~\ref{fig.rates};
model FS1 and FS2).  The corresponding detection numbers for advanced
LIGO full sensitivity could be non-zero (for model FS1: $\sim 2$
detections per year of observations), but still remain very low 
relative to Pop~I/II BH-BH mergers ($\sim 700$; see Tab.~\ref{tab.det}).  
Additionally, as we pointed out in Section~\ref{sec.detection} Pop~III 
BH-BH mergers cannot be reliably distinguished from Pop~I/II BH-BH 
mergers by advanced LIGO.

\subsection{Analytical arguments for low Pop~III BH-BH merger rates}

The merger rate density at $z=0$ can be expressed as 
\begin{equation}
{\cal R}= \int dt {SFR(t_{today}-t) }  X_{BHBH} {dN\over dt}
\label{rate-def1}
\end{equation}
where $t$  is the cosmic time.
In the case of Pop~III the star formation is concentrated around very 
early times right after the Big Bang (see Fig.~\ref{fig.sfr}). 
The distribution of the delay times can be calculated using the delays 
computed numerically (see Fig.~\ref{fig.del} and Fig.~\ref{fig.delK}). 
In evaluating the integral~\ref{rate-def1} we see that the integrand defined 
by the star formation is concentrated  around $t_{form}=300$ Myr. Thus we can express 
it as
\begin{equation}
{\cal R}=  X_{BHBH} {dN\over dt}(t_{today}-t_{form}\approx 13500{\rm Myr})\int dt {SFR(t)
}
\label{rate-def2}
\end{equation}

Therefore the current, local  merger rate density of BH-BH binaries originating in 
Pop~III stars is due to the binaries that were formed in distant past and took 
about $13.1 - 13.6$ Gyr to merge. In such a case one can estimate the current merger 
rate density in the following way. Let us first calculate the total mass density 
in the Pop~III stars: 
\begin{equation}
\rho_{PopIII} = \int_0^{T_{Hubble}} SFR(z(t)) dt 
\end{equation}

\begin{figure}
   \hspace*{-0.5cm}
   \includegraphics[width=9.1cm]{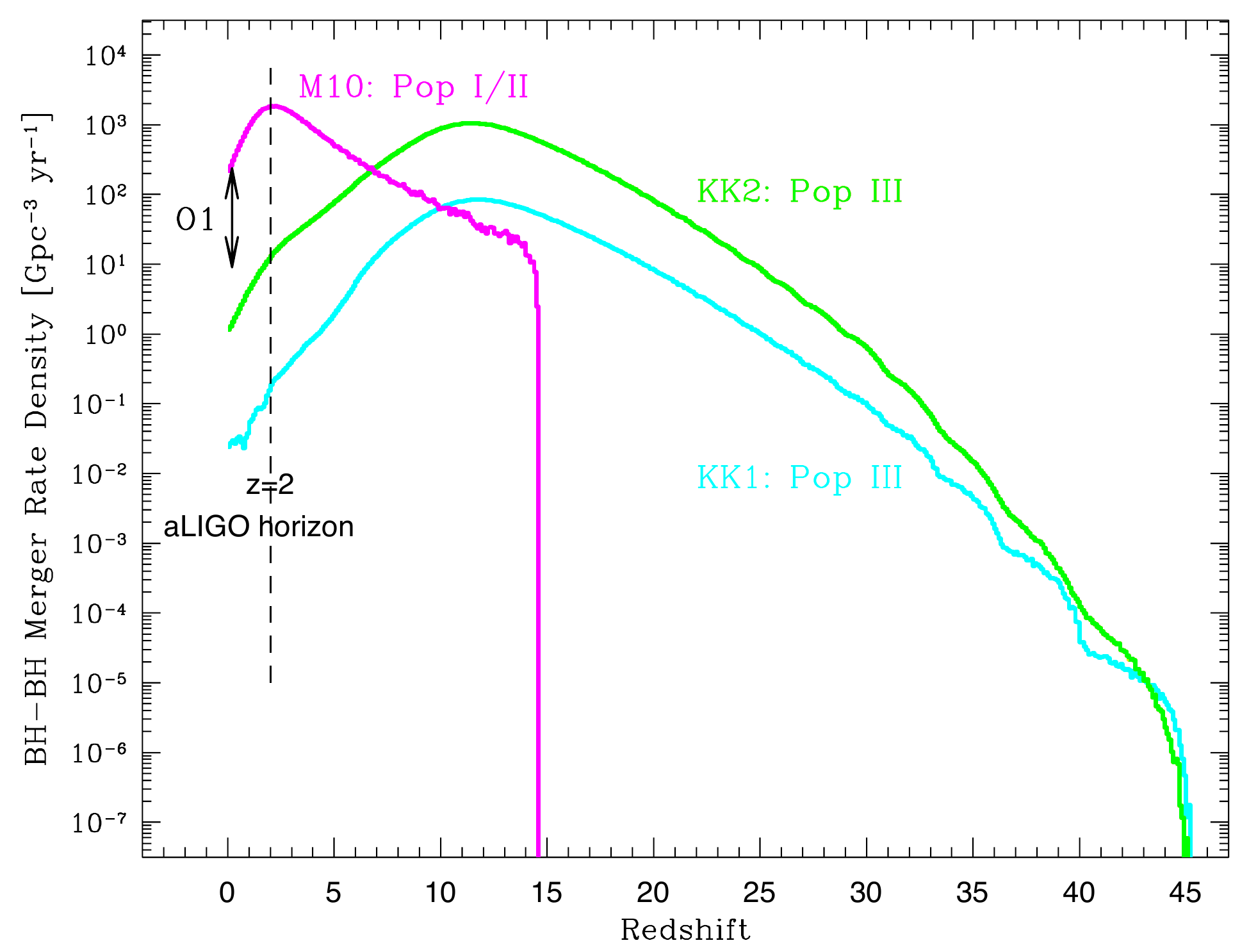}
   \vspace*{-0.0cm}
   \caption{
   Merger rate density for BH-BH binaries in a function of redshift for
   models KK1 and KK2. 
   For comparison we also show merger rate density for Pop I/II stars:
   model M10. The local BH-BH merger rate density measured by advanced 
   LIGO during O1 observations is marked. We also mark advanced LIGO 
   horizon for its full (design) sensitivity. 
   }
   \label{fig.ratesK}
   \end{figure}

\begin{figure}
   \hspace*{-0.5cm}
   \includegraphics[width=9.1cm]{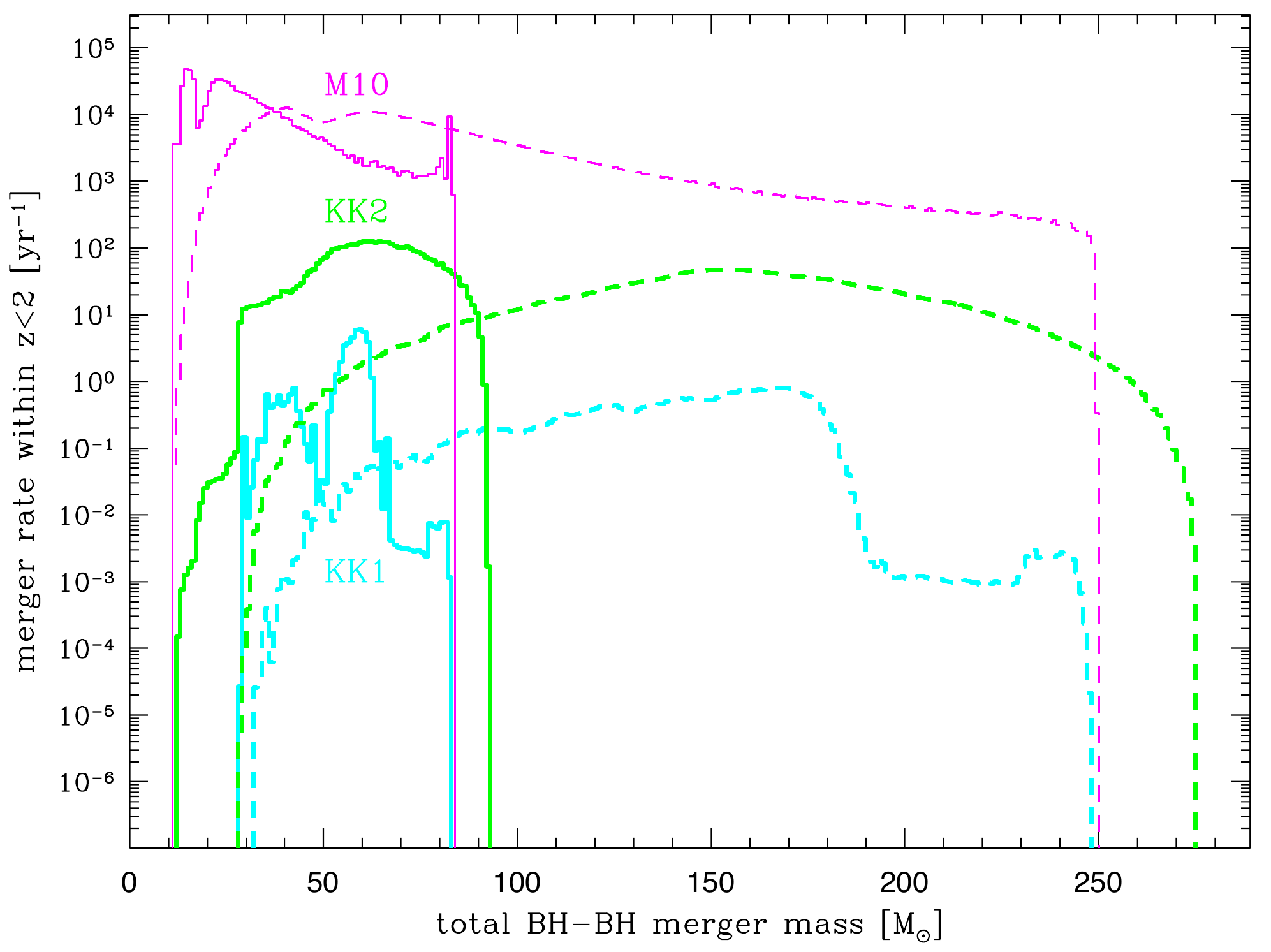}
   \vspace*{-0.0cm}
   \caption{
   Total intrinsic (solid lines) and redshifted (dashed lines) BH-BH merger mass 
   for BH-BH  mergers that take place within redshift of 2 ($z<2$) for
   models KK1 and KK2. Average mass for these models is $M_{\rm tot}=55.8\msun$ 
   ($M_{\rm tot,z}=141\msun$) for model KK1 and $M_{\rm tot}=62.8\msun$ 
   ($M_{\rm tot,z}=158\msun$) for model KK2. 
   For comparison Pop~I/II BH-BH mergers (model M10) have average mass: 
   $M_{\rm tot}=29.7\msun$ ($M_{\rm tot,z}=73.7\msun$). 
   }
   \label{fig.totmassK}
   \end{figure}

\begin{figure}
   \hspace*{-0.5cm}
   \includegraphics[width=9.1cm]{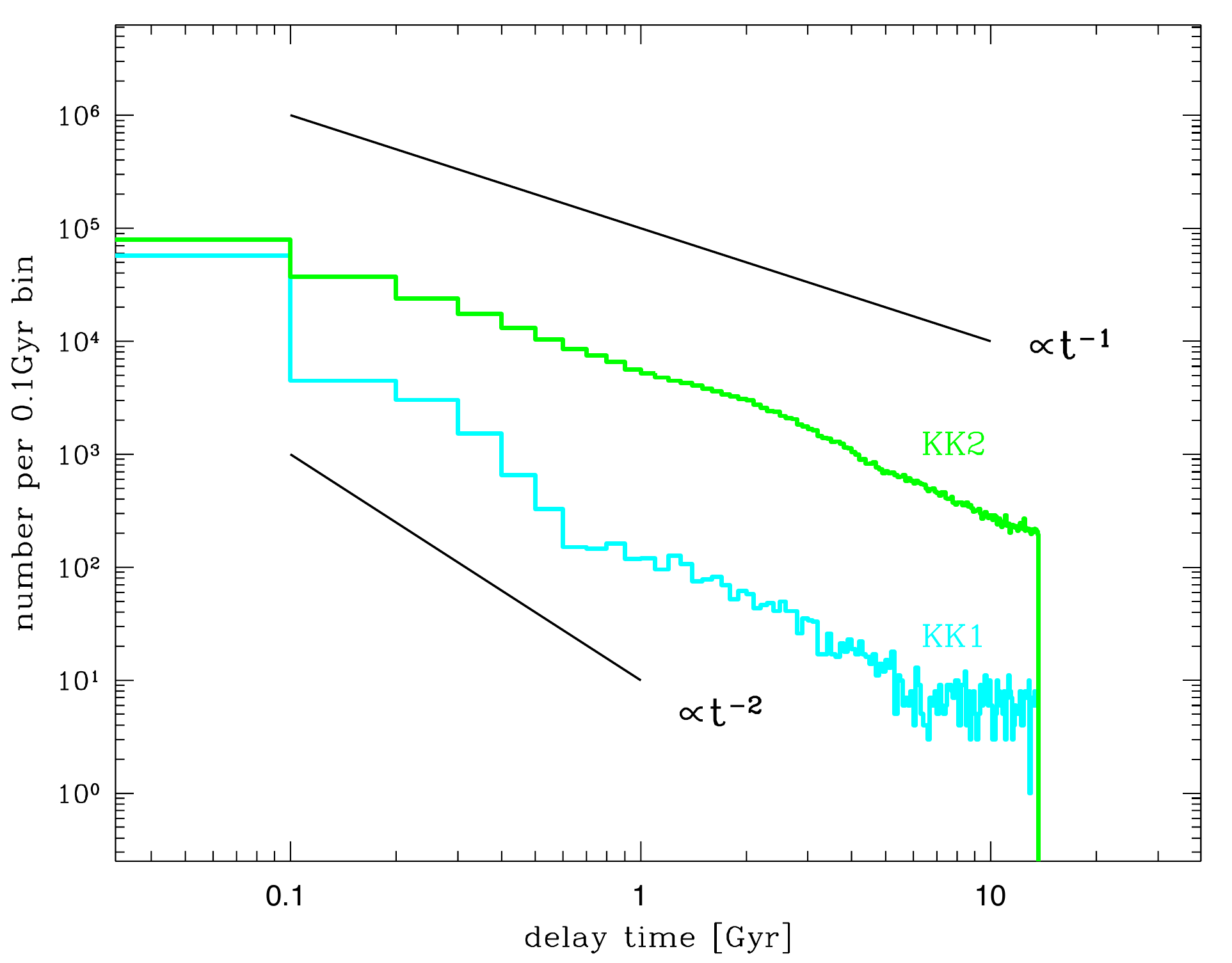}
   \vspace*{-0.0cm}
   \caption{
   Delay times for BH-BH mergers for models KK1 and KK2. It is
   clearly seen that the delay times (from star formation to the merger) 
   are short to intermediate; the average delay time is $192$ Myr for model
   KK1 and $1,580$ Myr for model KK2. 
   }
   \label{fig.delK}
   \end{figure}

Integrating the star formation rate presented in Figure~\ref{fig.sfr} we obtain 
$\rho_{PopIII} = 8\times 10^6\,M_\odot{\rm Mpc}^{-3}$. The value of the delay 
time distribution at $_t{del}=t_{today}-t_{form} \approx 13.5$ Gyr can be easily 
read off Figure~\ref{fig.delK} 
with the use of Table~\ref{tab.eff} as ${dN\over dt} |_{del} = n_i/(10^8\times N)$, where 
$n_i$ is the number of binaries in a $10^8$ yr bin at $\approx 13.5$ Gyr as shown in 
Figure~\ref{fig.delK}, and $N$ is the number of merging BH-BH binaries in the simulation 
from column 4 of Table~\ref{tab.eff}. 
We obtain ${dN\over dt} |_{del} \approx 7\times 10^{-13}$yr$^{-1}$ for KK1, 
and ${dN\over dt} |_{today} \approx 2\times 10^{-12}$yr$^{-1}$ for KK2.
Using the values of BH-BH merger formation efficiencies of $3.5\times 10^{-5}$ for KK1, and 
$5.8\times 10^{-4}$ for KK2,  we obtain the local merger rate densities for the model 
KK1 and KK2: $R_{KK1} \approx 0.02$ Gpc$^{-3}$yr$^{-1}$, and 
$R_{KK2} \approx 1.1$ Gpc$^{-3}$yr$^{-1}$, quite close to the result of the detailed 
calculation presented in Figure~\ref{fig.ratesK}.

This represents the fact that the local merger rate density is a result of the 
tail of the distribution of the delayed mergers produced at $z\approx10$. Moreover the 
BH-BH production efficiency in Pop~III stars is similar to the one of very low 
metallicity stars. This is mainly due to the fact that the Pop~III IMF starts  
at $10\msun$, and therefore a large fraction of Pop~III stars leads to formation 
of BHs. However the total mass of stars in Pop~III is much smaller than the total 
mass of stars in Pop~I/II. Thus the local merger rate density of Pop~III mergers is 
suppressed in comparison to the Pop~I/II stars for two reasons: smaller total mass 
of stars in this population, and the fact that the Pop~III star formation has 
ceased roughly $10$ Gyr ago ($z\approx2$).  

The local merger rate density of Pop~III BH-BH mergers could  be increased by 
increasing the BH-BH merger formation efficiency or by altering the delay time 
distribution. The highest the BH-BH formation efficiency that can ever be obtained 
is when all the BHs formed end up in merging BH-BH binaries. For a binary fraction 
of $f_{bin}=1/3$ and with the KK1 IMF we obtain the maximum effciency of BH-BH
merger formation of $7.5\times 10^{-3}$, which means that BH-BH mergers are 
produced with the efficiency of $\approx 0.5$\% of the maximal possible value. 
For model KK2 production of BH-BH proceeds at $\approx 6\%$ of the maximum value.
The delay distribution can not be changed by much without tuning the initial 
distribution of orbital separation especially to do so.

The merger rate density of Pop~I/II BH-BH binaries cannot be that easily calculated. 
The major contributions to the integral of equation~\ref{rate-def1}
 comes from two regions. The first is the recent 
star formation with short delay times, and the second are the long delay binaries for 
which the delay distribution factor is small but the SFR contributed from a long stretch 
of time. The two contributions are roughly of the same order of magnitude. Thus the 
merger rate density is approximately, but not exactly, proportional to the SFR.

Using the similar arguments as above we can estimate the maximum BH-BH formation 
efficiency for Pop~I/II stars - assuming Salpeter IMF (with $\alpha=-2.3$), 
$f_{\rm bi}=1/2$ binary fraction and ignoring the effects of binary evolution. With 
these assumptions we obtain the maximum BH-BH formation efficiency of $4\times 10^{-4}$. 
Thus, for model M10 with the metallicity of 0.0002, the actual BH-BH formation efficiency is 
$3.0\times10^{-5}$ (see Tab.~\ref{tab.eff}) which is about  $10\%$ of the maximum value.
The overall efficiency is a mean value coming from averaging over $32$ metallicities 
in model M10.

\section{Conclusions}

A number of recent studies exploring BH-BH and BH-NS mergers from Pop~III
stars have been based on evolution specific to metal free stars, but
initial conditions employed in these studies were not relevant for
Pop~III stars~\citep{Kinugawa2014,Kinugawa2016a,Kinugawa2016b}.  On
the other hand, in another recent study, \cite{Hartwig2016} employed
more appropriate physical initial conditions for Pop~III stars, albeit
the results were based on evolutionary calculations relevant for Pop~I
stars.  

Ours is the first computation of BH-BH merger rates which
simultaneously combines initial evolutionary conditions specific to
Pop~III stars, as derived from $N$-body simulations of binary
formation in primordial halos, with binary evolution models
for metal-free stars. All of the studies listed above are important
steps in the process of understanding the formation of double compact
objects from the first stars. The work is still in progress and many
open questions about the formation and physical properties of Pop~III
stars remain open. In the following we present our limited set of
conclusions based on the numerical simulations of formation and
evolution specific to the first stars. 

The initial conditions in our study were derived from cosmological, 
hydrodynamical simulations of dark matter halos at high redshifts, and include 
the initial mass function, binary fractions, separations, mass ratios and
eccentricities determined via N-body simulations.
The stellar evolution model updates pre-existing
models to take into account limited radial expansion and lack of
stellar wind mass loss from Pop~III stars, and produces a
realistic spectrum of Pop~III black holes. We have also adopted
a very optimistic star formation rate for Pop~III stars, that
most likely places an upper limit on the number of these stars in
Universe. Therefore our conclusions, in the context of Pop~III
BH-BH, BH-NS and NS-NS merger rates and chances of their detections, are 
most likely on the high side. These are our basic conclusions.

\begin{enumerate}
	
	\item
	The initial conditions for the evolution of Pop~III binaries are very 
	different from those of Pop~I/II binaries. In fact, no Pop~I/II initial
	distribution should be applied in studies of Pop~III binaries. 
	For example, the Pop~III initial mass function increases with star mass
	(increasing massive BH formation) instead of the steep power-law fall off 
	observed for Pop~I/II stars (see Fig.~\ref{fig.imf}). 
	Additionally, the initial orbital separations for Pop~III binaries are 
        depleted at separations typical for BH-BH merger formation $\sim 1000\rsun$  
	(limiting BH-BH merger formation) as compared with Pop~I/II binaries 
        (see Fig.~\ref{fig.a0}).
	
	\item 
	We have considered, depending on the model, star formations in
        the Universe specific to Pop~I/II stars and Pop~III
        stars. Since we have adopted rather pessimistic estimates for
        the Pop~I/II formation rate and a rather optimistic one for
        the Pop~III formation rate (see Fig.~\ref{fig.sfr}), our
        results should be viewed as delivering an optimistic ratio of
        Pop~III to Pop~I/II BH-BH mergers.  Despite this fact, only
        $\sim 0.3\%$ of all stars in our simulation are Pop~III, while
        the rest are Pop~I/II.  This stark contrast makes it extremely
        difficult for Pop~III BH-BH mergers to make a large
        contribution to the LIGO event rate.
	
	\item 
	We have estimated the formation efficiency of BH-BH mergers from Pop~III
	binaries to be significantly, but not overwhelmingly, higher than that from 
	Pop~I/II binaries. This efficiency is a strong function of metallicity 
	for Pop~I/II stars. Pop~III stars form BH-BH mergers (per unit
	of star forming mass) a factor of $\sim$~3, 10, 500 times more effectively than Pop~I/II
	binaries at metallicity of $Z=0.0002, 0.002, 0.02$, respectively (see
	Tab.~\ref{tab.eff}). 
	
	\item 
	The Pop~III binaries in the framework of our model, in which we do
	not consider stars with very rapid rotation, form along classical binary
	evolution channels that involve a common envelope phase (see Tab.~\ref{tab.evol}).
	The combination of initial distribution of orbital separations, with CE orbital
	decay followed by gravitational radiation-induced inspiral,
	generates a steep delay-time distribution of Pop~III BH-BH mergers
	$\propto t^{-3} - t^{-1}$ (see Fig.~\ref{fig.del}).
	Given that Pop~III stars form at high redshifts and evolve toward BH-BH 
        mergers with relatively short delay times, they are unlikely to make a 
        significant contribution to the LIGO signal. 
	
	\item
	The combination of all our calculations leads to a full cosmological
	prediction of the merger rate density and the detection chances of Pop~III
	BH-BH mergers. The merger rate density of Pop~III BH-BH systems is found
	at the level of $0.1\gpy$ (see Fig.~\ref{fig.rates}) within the reach of
	advanced LIGO at its full projected design sensitivity ($z<2$).
	For comparison, recent LIGO observations (O1) have constrained the local BH-BH
	merger rate ($z<0.1 - 0.2$) to be $R_{\rm BHBH}=9 - 240\gpy$~\citep{LigoO1b}.
	Our results strongly indicate that Pop~III BH-BH mergers cannot
	comprise a significant share of existing and future advanced LIGO detections.
	However, our predicted detection rate for Pop~III BH-BH mergers is not zero, 
	and LIGO at its full design sensitivity may potentially detect a couple of such
	mergers per year (see Tab.~\ref{tab.det}). This needs to be contrasted with 
	predictions of tens-to-hundreds of BH-BH mergers a year from Pop~I/II 
	stars \citep[e.g.,][]{Belczynski2016b,deMink2016,Rodriguez2016b}.

        \item
        The identification of a Pop~III origin of a BH-BH merger event is not 
        straightforward, as it lacks a smoking gun signature: BH masses alone are 
        not likely to distinguish Pop~III from Pop~I/II mergers in advanced LIGO 
        era (see Sec.~\ref{sec.detection}). However, with 3rd generation 
        interferometers it may be possible to see a peak of the merger rate 
        distribution at $z\gtrsim 10-12$ which may be attributed to Pop~III stars 
        (see Fig.~\ref{fig.rates}). 
        It is also possible that stochastic gravitational wave background may 
        provide means to identify Pop~III BH-BH mergers that are forming with
        heavier (on average) mass and very different redshift distribution than 
        Pop~I/II BH-BH mergers. It remains to be tested whether Pop~III BH-BH 
        mergers that are formed in our simulations could be detected and identified 
        by AdLIGO at its full sensitivity through stochastic gravitational wave 
        background\footnote{Data on Pop~III mergers can be found at 
        \url{http://www.syntheticuniverse.org} and any extra data is available upon
        request from Chris Belczynski.}.  
        We note that any Pop~III BH-BH merger identification would carry profound 
        implications for our understanding of the environments in which the first 
        stars in the Universe  formed. In particular, we have shown how, if Pop~III 
        stars generally form as a result of fragmentation of multiple, less massive 
        protostars around a more massive one on scales of tens of AU, as found e.g. 
        by the simulations of \citet{Greif+12}, then no Pop~III merger should ever 
        be detected, even by future gravitational wave detectors. However, even a 
        single event would argue towards formation in halos with spatial scales of 
        several thousands of AU \citep{StacyBromm13}.

\end{enumerate}

\vspace{0.5cm}

{\bf Acknowledgments} 
Authors would like to thank Kohei Inayoshi, Ilya Mandel and Salvatore Vitale 
for useful comments. 
KB acknowledges support from the Polish National Science Center (NCN) grants: 
Sonata Bis 2 (DEC-2012/07/E/ST9/01360), OPUS (2015/19/B/ST9/01099), OPUS 
(2015/19/B/ST9/03188) and MAESTRO (2015/18/A/ST9/00746). 
Results in this paper were obtained using the high-performance LIred 
computing system at the Institute for Advanced Computational Science at Stony 
Brook University, which was obtained through the Empire State Development 
grant NYS \#28451.
RP acknowledges support from the NSF under grant AST-1616157. 
EB was supported by NSF Grant No. PHY-1607130 and by FCT contract
IF/00797/2014/CP1214/CT0012 under the IF2014 Programme. This work was
supported by the H2020-MSCA-RISE-2015 Grant No. StronGrHEP-690904.
TB was supported by the NCN grant 2014/15/Z/ST9/00038.

\bibliography{biblio}

\end{document}